\documentclass[preprint,review,12pt]{elsarticle}
\usepackage{amssymb}
\usepackage{float}
\usepackage{subfig}

\journal{Astroparticle Physics}

\begin{document}

\begin{frontmatter}

\title{ CORSIKA Simulation of the Telescope Array Surface Detector }

\author[]{\small \par\noindent
T.~Abu-Zayyad$^{26}$,
R.~Aida$^{27}$,
M.~Allen$^{26}$,
R.~Anderson$^{26}$,
R.~Azuma$^{20}$,
E.~Barcikowski$^{26}$,
J.W.~Belz$^{26}$,
D.R.~Bergman$^{26}$,
S.A.~Blake$^{26}$,
R.~Cady$^{26}$
% ...

\par\noindent
{\footnotesize\it
% ...
$^{20}$ Tokyo Institute of Technology, Meguro, Tokyo, Japan \\
% ...
$^{26}$ University of Utah, Department of Physics \& Astronomy and High Energy Astrophysics Institute, Salt Lake City, Utah, USA \\
$^{27}$ University of Yamanashi, Interdisciplinary Graduate School of Medicine and Engineering, Kofu, Yamanashi, Japan \\
% ...
}
\par\noindent
}

\begin{abstract} The Telescope Array is the largest experiment studying 
ultra-high energy cosmic rays in the northern hemisphere.  The detection area 
of the experiment consists of an array of 507 surface detectors, and a 
fluorescence detector divided into three sites at the periphery.  The viewing 
directions of the 38 fluorescence telescopes point over the air space above 
the surface array.  In this paper, we describe a technique that we have 
developed for simulating the response of the array of surface detectors of the Telescope Array experiment.  The two primary components of this method
are (a) the generation of a detailed CORSIKA Monte Carlo simulation with all 
known characteristics of the data, and (b) the validation of the simulation 
by a direct comparison with the Telescope Array surface detector data. This 
technique allows us to make a very accurate calculation of the acceptance of 
the array.  We also describe a study of systematic uncertainties in this acceptance 
calculation. 
\end{abstract}

\begin{keyword}
cosmic ray \sep extensive air shower \sep simulation \sep surface detector 

\end{keyword}

\end{frontmatter}

\section{Introduction}

The Telescope Array (TA) is the largest experiment studying ultrahigh energy cosmic rays in the 
northern hemisphere.  It is located in Millard County, Utah, and consists 
of a surface detector (SD) of 507 scintillation counters, each of area 3m$^2$, deployed in a 
grid of 1.2 km spacing, plus a set of 38 fluorescence telescopes located at 
three sites around the SD looking inward over the array.  Both detector 
systems of TA started collecting data in 2008.  

Measurements of the differential flux of cosmic rays, as a function of 
energy, have historically played an important role in the study of ultra-high energy 
cosmic rays (UHECRs).  Foremost among these is the high energy break in 
the spectrum at $5 \times 10^{19}$ eV, called the GZK 
cutoff~\cite{greisen}\cite{zk}\cite{hiresmono:gzk}\cite{auger_gzk}, which 
provides convincing evidence for the extra-galactic origin of the highest 
energy cosmic rays. 

An important experimental technique used in the 
spectrum measurement is the calculation, using the Monte Carlo simulation 
method, of the efficiency with which the detector observes cosmic ray 
induced extensive air showers.  Prior to the TA 
experiment, high-fidelity Monte Carlo (MC) simulations have been available 
for fluorescence detectors (FDs), which measure the fluorescence light 
emitted by nitrogen molecules excited by the passage of shower particles 
in their vicinity. Accurate simulations for the other major detector type, 
surface scintillation arrays, have only recently become possible with 
the rapid growth of computational and storage capacity over the past 
decade, coupled with the maturity of sophisticated and realistic  shower 
generation codes over the same time frame. In particular, the difficulty 
of generating accurate Monte Carlo simulations of air showers has limited 
the surface array technique to the energy regime where the detector is 
100\% efficient \cite{auger:optdist}; i.e., only at the high energy end 
of the detector's sensitive range.  

In order to simulate accurately the ground-level particle densities 
measured by surface detectors, along with their fluctuations,  a shower 
generator code needs in principle to track every particle created in the 
avalanche process down to below its critical energy.  In practice, 
available CPU power and storage space limit one to generating only a 
small number of shower particles, insufficient for an accurate calculation of 
detector acceptance, or for a useful comparison of data and MC 
distributions.  An approximation technique called "thinning" 
\cite{Hillas:1997tf} typically is used in programs like CORSIKA~\cite{corsika} 
and AIRES~\cite{aires} 
to reduce CPU time requirements.  Under the thinning approximation,  
nearly all  particles with energies below a preselected threshold (orders 
of magnitude higher than the critical energy) are removed from the shower.  
Only a few representative particles are kept with weights to account for 
those, in the same region of phase space, that have been "thinned" out. 

The thinning method usually gives an adequate description of particle 
distributions in the core region of a shower where enormous numbers of 
particles are found (and where essentially all of the fluorescence light 
is generated).  For surface detectors, which sample the particle density 
at ground level, the enormous flux saturates any counter in proximity to 
the shower core. Typically, useful sampling is based on detectors at the 
scale of the detector spacing or more.  For experiments, like TA, that 
are optimized to measure the highest energy cosmic rays, this distance 
scale is of the order of a kilometer. While a thinned shower is able to 
reproduce the average particle densities reasonably well on the kilometer 
scale from the shower core, the weighted particles cannot model the 
shower-to-shower fluctuations or even the fluctuations at different 
azimuthal angles around the shower core.   The RMS deviations from the 
average densities in a thinned shower are typically off by an order of 
magnitude or more from that obtained from those seen in the few "unthinned" 
showers one can afford to generate. Thinning is therefore too crude of an 
approximation to give a faithful representation of even the simulated air 
shower itself, let alone real cosmic-ray induced showers. Some experiments 
have claimed to overcome this intrinsic difficulty by restricting their 
analysis to the highest energy range where the efficiency of the detector 
approaches unity.  However, if quality cuts are used to select only a 
subset of the data, then the use of a simulation is still needed to 
calculate acceptance. In that case the use of thinning can and probably 
does introduce significant systematic biases because the thinned Monte 
Carlo (MC) simulation cannot accurately reproduce the tails expected in 
the distribution of cut parameters.  Quality cuts are invariably used to 
remove outliers in such tails.

In the simulation of air showers for calculating the acceptance of the 
Telescope Array experiment, we have developed a "de-thinning" procedure 
to compensate for the shortcomings of the thinning.  Using the thinned 
CORSIKA output, we replace each representative particle of weight $w$ 
with an ensemble of $w$ particles propagated in a cone about the weighted 
particle. A detailed prescription of our de-thinning process was published 
in an earlier article~\cite{dethinned}.  In that article, careful 
comparisons were made between de-thinned and non-thinned showers (the latter 
referring to showers generated without any thinning), and excellent 
agreement was found in the statistical properties of the two sets of 
simulations.  Our de-thinned sample overcame all of the essential 
shortcomings of the thinning approximation. 

In this paper, we describe 
the actual application of the de-thinning process to the simulation of 
the Telescope Array experiment. Detailed comparisons are shown for key 
distributions (those that directly affect the acceptance calculation) 
between TA data and the de-thinned MC shower sample. The excellent 
agreement in these comparisons serve to demonstrate the high degree of 
accuracy of the simulation in reproducing the properties both of the 
detector and of the data.

This paper is the last in a series of three describing simulation 
techniques used for the surface detector of the Telescope Array 
experiment~\cite{dethinned}\cite{parallel}.  Sections~2 and 3 give an overview 
of the TA surface detector and its data analysis. Section~4 describes the process 
of generating de-thinned CORSIKA showers for the TA SD, with events 
generated according to  previous measurements of the UHECR spectrum and 
composition, and including a detailed simulation of each scintillation 
counter.  Validating the Monte Carlo simulation is described in Section~5, and the
experimental resolution is presented in Section~6.  Determining the energy scale
using events seen by both the fluorescence and surface detectors is given in Section~7,
and a study of systematic uncertainties and biases is described in Section~8.

\section {TA Surface Detector Data}

\begin{figure}[t,b]
  \begin{center}
    \includegraphics[width=0.5\textwidth]{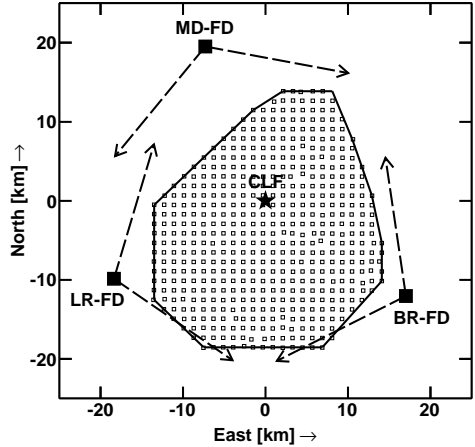}
  \end{center}
  \caption{The physical layout of the Telescope Array.  The surface detectors
    are represented by small open squares.  Additionally the positions and
    fields of view are shown for all three air-fluorescence stations.}
  \label{fig:layout}
\end{figure}

The TA surface detector has been described previously~\cite{ta:1}\cite{ta:2}\cite{sdspec}.  
In Figure~\ref{fig:layout},
we see the physical layout of all components of TA.  Each SD counter consists of two layers of plastic scintillator, 3 m$^2$ in area, and read out independently by two photomultiplier tubes.  Scintillation light is guided to the photomultiplier tubes by a system of wavelength-shifting fibers set in grooves in the scintillator.  These counters are calibrated every 10 minutes~\cite{tasd:performance} using a histogram of pulse heights recorded for events triggering both layers of scintillators in time coincidence.  The resulting distributions typically consist of a peak at low integrated pulse area, accompanied by a tail toward higher pulse area.  The peak itself corresponds to the signal from single muons passing through both scintillators, and the centroid of the peak then defines the average signal for a minimum-ionizing particle (MIP) for that channel.

The waveforms from the two scintillators are sampled by a 50 MHz FADC system [14].  A real time integration process is used to trigger each counter: waveforms of pulses with integrated areas corresponding to at least 1/3 MIPs are saved with a corresponding GPS time stamp.  The detection of a pulse of 3 MIPs or larger is reported to a central data acquisition system via the radio communication system.  A trigger for the TA SD occurs when three adjacent counters have energy deposits equivalent to at least three MIPs in each counter within an 8 $\mu$sec window.  When the SD trigger conditions are met, all counters in the array are polled.  Saved waveforms (of minimum 1/3 MIPs) with time stamps within 64 microseconds of the event trigger are read out over the radio link.  The pulse area contained in recorded waveforms, stored in raw FADC units,  are converted to units of vertical-equivalent muons (VEM).   A vertical minimum-ionizing muon deposits on average 2.05 MeV of ionization energy in each scintillator layer.  The conversion process uses the calibration histograms collected every 10 minutes, but also incorporates the simulated detector response to single muons and other secondary
particles produced in air showers induced by TeV cosmic rays. 

\begin{figure}[t,b]
  \begin{center}
    \includegraphics[width=0.5\textwidth]{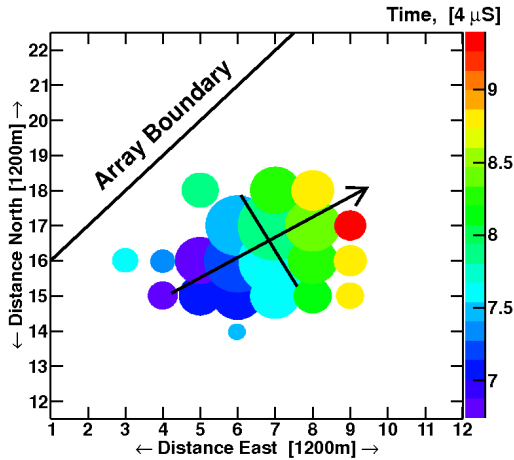}
  \end{center}
  \caption { A typical high energy event seen by the TA SD. Each
    circle represents a counter that participated in the event.  The area of
    each circle is proportional to the logarithm of the VEM signal size for that counter.  The measured arrival time of the shower at each counter is denoted by the color of the circle.  
    The arrow
    represents the projection of the shower axis onto the ground, $\hat{u}$, 
    and the intersection between this arrow and its perpendicular bisector marks the location of the shower core. }
  \label{figure:event_display}
\end{figure}

The application of the calibration and the signal analysis extract the following information from each counter participating in the event: (a) An integrated particle count in units of VEM, (b) the arrival time of the shower, and (c) the spatial coordinates of the counter.  These quantities are then used to reconstruct the shower trajectory and the energy of the primary cosmic ray.  Figure~\ref{figure:event_display} 
shows a footprint of a
typical high energy event.

\section {Event Reconstruction and Selection Cuts}

The event reconstruction procedures used for TA SD data are based on parametrizations 
and procedures originally developed by the AGASA Collaboration~\cite{Takeda:2002at},  modified
 to match the characteristics of the TA detectors~\cite{tasd_ichep2010}.

\begin{figure*}[t,b]
  \centering
  \subfloat[]{\includegraphics[width=0.5\textwidth]{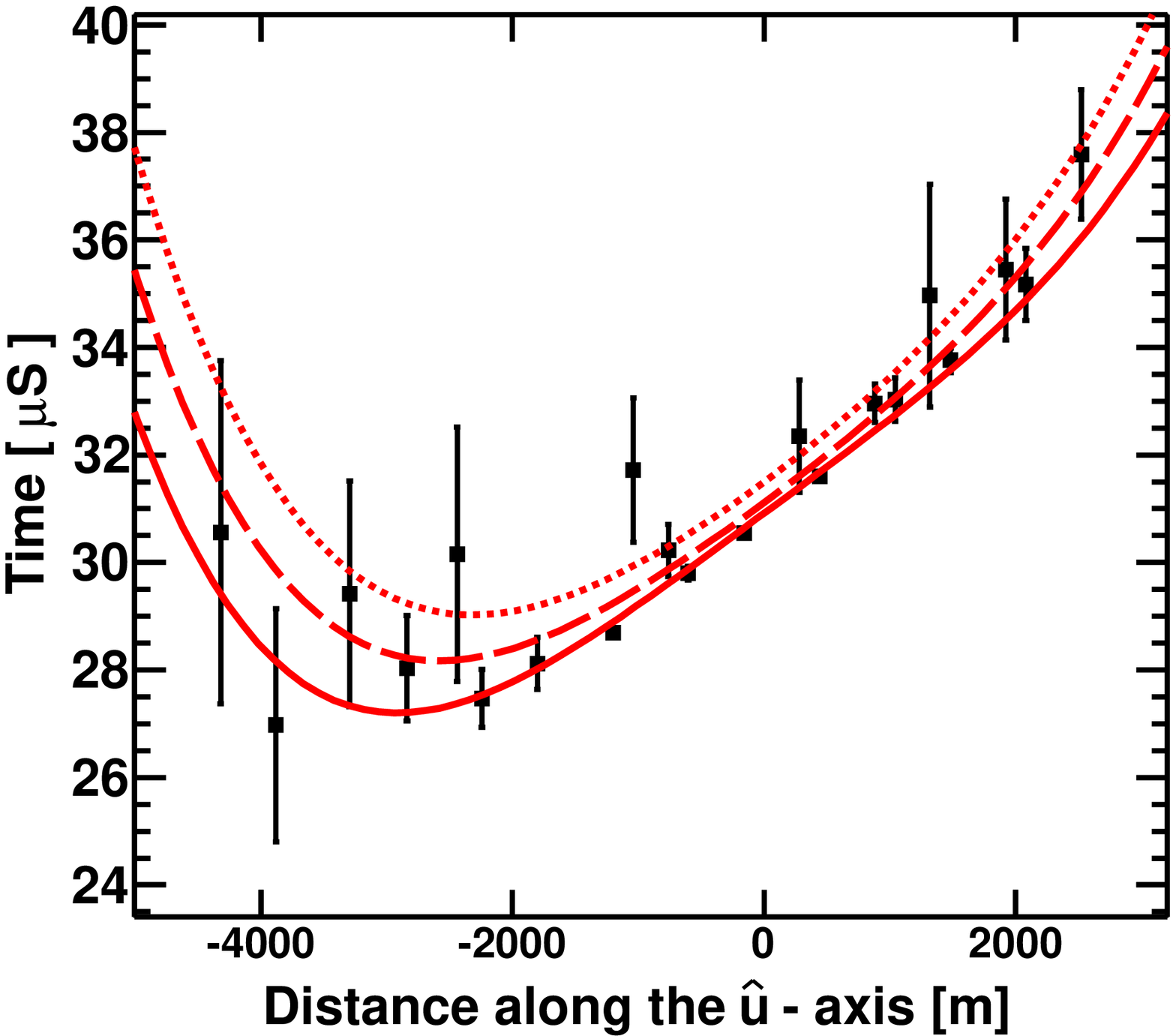}\label{figure:timefit}}
  \subfloat[]{\includegraphics[width=0.5\textwidth]{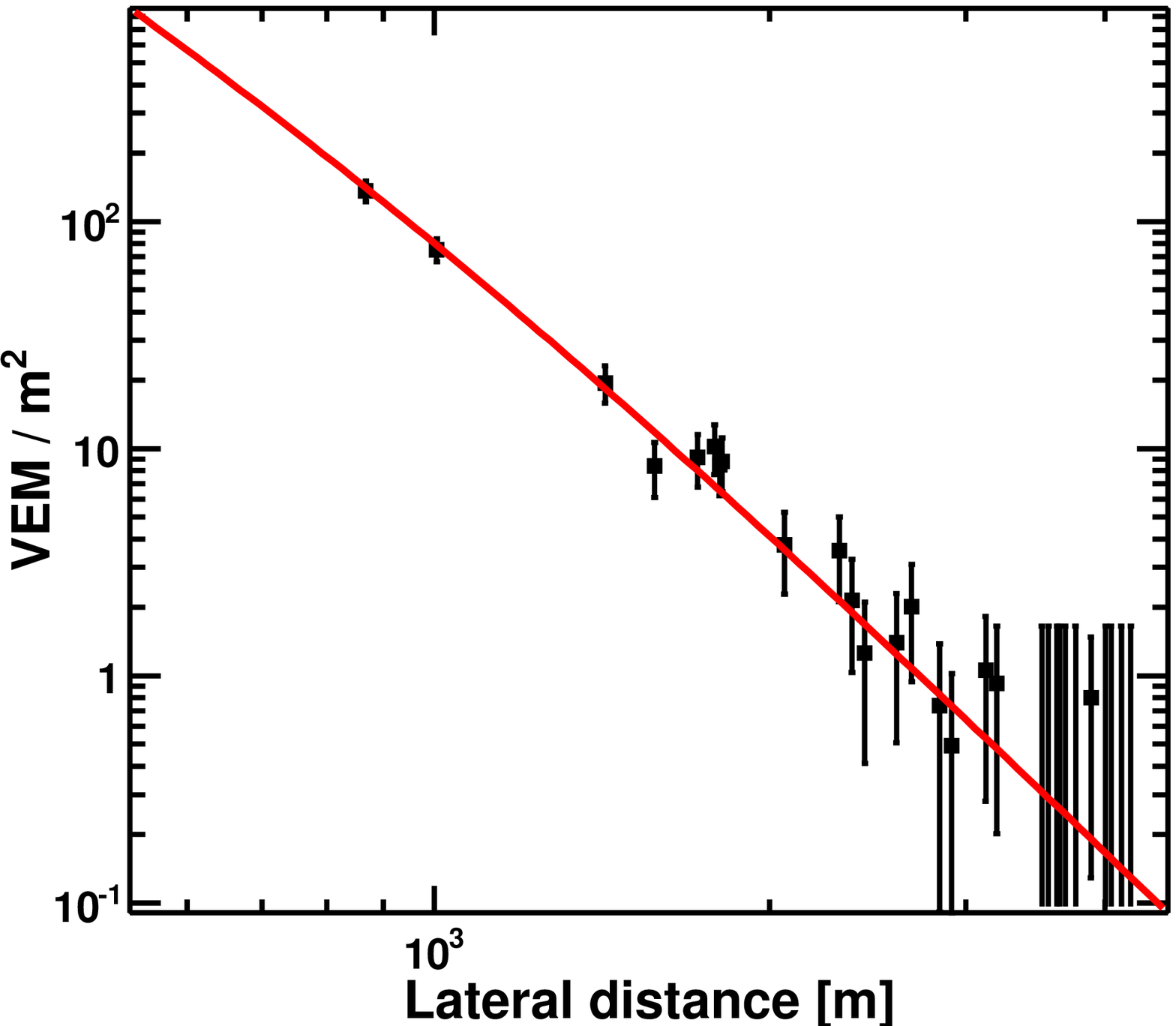}\label{figure:ldffit}}
  \caption{ Time and lateral distribution fits for a typical TA SD
    event. \protect\subref{figure:timefit} Counter time versus distance from the
    shower core along the $\hat{u}$-direction.  Points with error bars are
    the measured counter times.  The solid curve gives the times predicted by the modified
    AGASA fit for the counters lying on the $\hat{u}$-axis.  The dashed and
    dotted lines are the fit expectation times for the counters that are
    located 1.5 and 2.0 km off the
    $\hat{u}$-axis, respectively. \protect\subref{figure:ldffit} Measured lateral distribution fit to
    the AGASA LDF function. The vertical axis is the signal density and the
    horizontal axis is the lateral/perpendicular distance to the shower core.  Event
    $S800$ is determined from the fit curve. }
  \label{figure:typical_event}
\end{figure*}

First, the shower axis is determined from the arrival times in the triggered counters.  These are fit to the AGASA-modified Linsley time delay function~\cite{Linsley:Td}\cite{agasa:timefit}.
Figure~\ref{figure:timefit}, 
shows this fit for a typical TA SD event.  For this work, the parameters of the original AGASA time delay function were adjusted to fit the overall characteristics of the TA SD data set, by considering the distributions of fit residual in distance-to-core, VEM signal sizes, and in zenith angle.
It should be emphasized that the adjustments were made based exclusively on actual TA SD data without any additional information from simulations, and are therefore model-independent.

The primary energy estimation of TA SD events is established by first 
measuring the charge density at 800m in lateral (perpendicular) distance from the shower axis 
($S800$)~\cite{auger:optdist}.  
The measured particle densities from the counters are fit to the modified AGASA lateral distribution
function (LDF)~\cite{agasa:results}, as shown in
Figure~\ref{figure:ldffit}.  The value at 800m, denoted as $S800$, is interpolated from this fit. 

In order to achieve reasonable detector resolutions in energy and pointing direction, but without losing an unreasonable fraction of events, we chose the following event selection cuts (both pre- and post- reconstruction).  These same cuts are applied to both data and Monte-Carlo in
the present TA SD analysis:

\begin{enumerate}
\item $N_{\mathrm{SD}} \ge 5$. At least 5 good counters per event.
\item $\theta < 45^{\circ}$.  Zenith angle less than 45 degrees.
\item $D_{\mathrm{Border}} \ge 1200$~m. Core position is within the array and 
  at least 1200~m away from the edge of the array.
\item $\chi_{\mathrm{G}}^{2}/\mathrm{d.o.f.} < 4$ and
  $\chi_{\mathrm{LDF}}^{2}/\mathrm{d.o.f.} < 4$.  Reduced values of
  $\chi^{2}$ of geometry ($\chi_{\mathrm{G}}^{2}/\mathrm{d.o.f.}$) and LDF
  ($\chi_{\mathrm{LDF}}^{2}/\mathrm{d.o.f.}$) fits are less than 4.
\item $\sqrt{\sigma_{\theta}^{2} + \mathrm{sin}^{2}\theta \,
    \sigma_{\phi}^{2}} < 5^{\circ}$. Pointing direction uncertainty is
  less than 5 degrees. $\sigma_{\theta}$ and $\sigma_{\phi}$ are the
  uncertainties on zenith and azimuthal angles from the
  geometry fit.
\item $\sigma_{S800} / S800 < 0.25$.  Fractional
  uncertainty of $S800$ determination (from the LDF fit) is within 25\%.
\end{enumerate}

Table~\ref{table:tasd_cut_eff} displays the efficiency (fraction of
events retained) when the quality cuts are applied, incrementally for
3 energy slices.
\begin{table}[!htbp]
  \begin{center}
    \begin{tabular}{|l|l|l|l|l|}
      \hline Quality cut & Efficiency, & Efficiency, & Efficiency, \\ 
      & $E > 10^{18}$~eV & $E > 10^{18.5}$~eV & $E > 10^{19}$~eV \\
      \hline $N_{\mathrm{SD}} \ge 5$ & 0.674 & 0.931 & 0.973 \\
      \hline $\theta < 45^{\circ}$ & 0.741 & 0.702 & 0.677 \\
      \hline $D_{\mathrm{Border}} \ge 1200$~m & 0.865 & 0.814 & 0.748 \\
      \hline $\chi_{\mathrm{G}}^{2}/\mathrm{d.o.f.} < 4, \, \chi_{\mathrm{LDF}}^{2}/\mathrm{d.o.f.} < 4$  
      & 0.928 & 0.938 & 0.981 \\
      \hline $(\sigma_{\theta}^{2} + \mathrm{sin}^{2}\theta \, \sigma_{\phi}^{2})^{1/2} < 5^{\circ}$ 
      & 0.656 & 0.925 & 0.995 \\
      \hline $\sigma_{S800} / S800 < 0.25$ & 0.534 & 0.887 & 0.995 \\
      \hline All cuts combined & 0.14 & 0.41 & 0.48 \\
      \hline
    \end{tabular}
  \end{center}
  \caption{Efficiency of the quality cuts}
  \label{table:tasd_cut_eff}
\end{table}

\section {Surface Detector Monte Carlo Simulation}

In simulating an air shower, the TA surface detector Monte Carlo uses the
CORSIKA 6.960~\cite{corsika} simulation package.  For the standard simulated event set, we selected the QGSJET-II-03~\cite{qgsjet} and FLUKA2008.3c~\cite{fluka1}\cite{fluka2} 
hadronic models for high and low energies, respectively.  For electromagnetic processes, the EGS4~\cite{egs4} electromagnetic model was used.  

The first step in generating a comprehensive simulation of the TA SD
data set is to create a library of thinned CORSIKA showers.  This library 
consists of 16,800 extensive air showers with primary energies distributed in 
$\Delta \log_{10}E=0.1$ bins between $10^{16.75}$~eV and $10^{20.55}$~eV.  The  number of showers in each bin ranges from 1000 in the lowest energy bin to 
250 in the highest energy bin.  These showers are simulated with 
zenith angles from $0^\circ$ to $60^\circ$ assuming an 
isotropic distribution. It is important to note that in our final analysis we only include events with $E>10^{18.0}$~eV and $\theta<45^\circ$. However, events must be simulated well beyond these limits in energy and inclination in order to give a complete understanding of our detector acceptance as well as our energy and angular resolutions.

\begin{figure*}[t,b]
  \centering
  \subfloat[]{\includegraphics[width=0.5\textwidth]{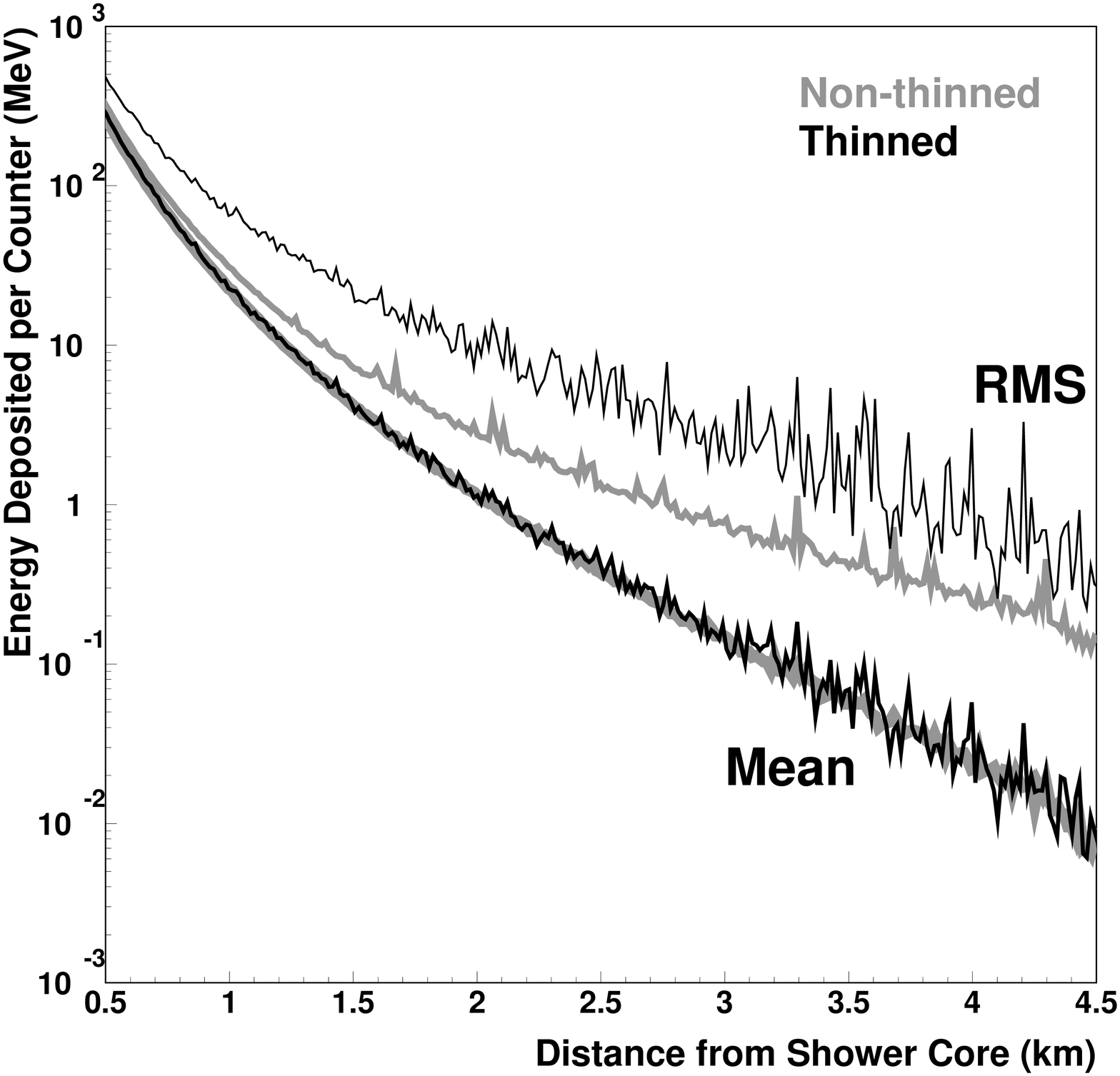}\label{fig:th_v_nth}}
  \subfloat[]{\includegraphics[width=0.5\textwidth]{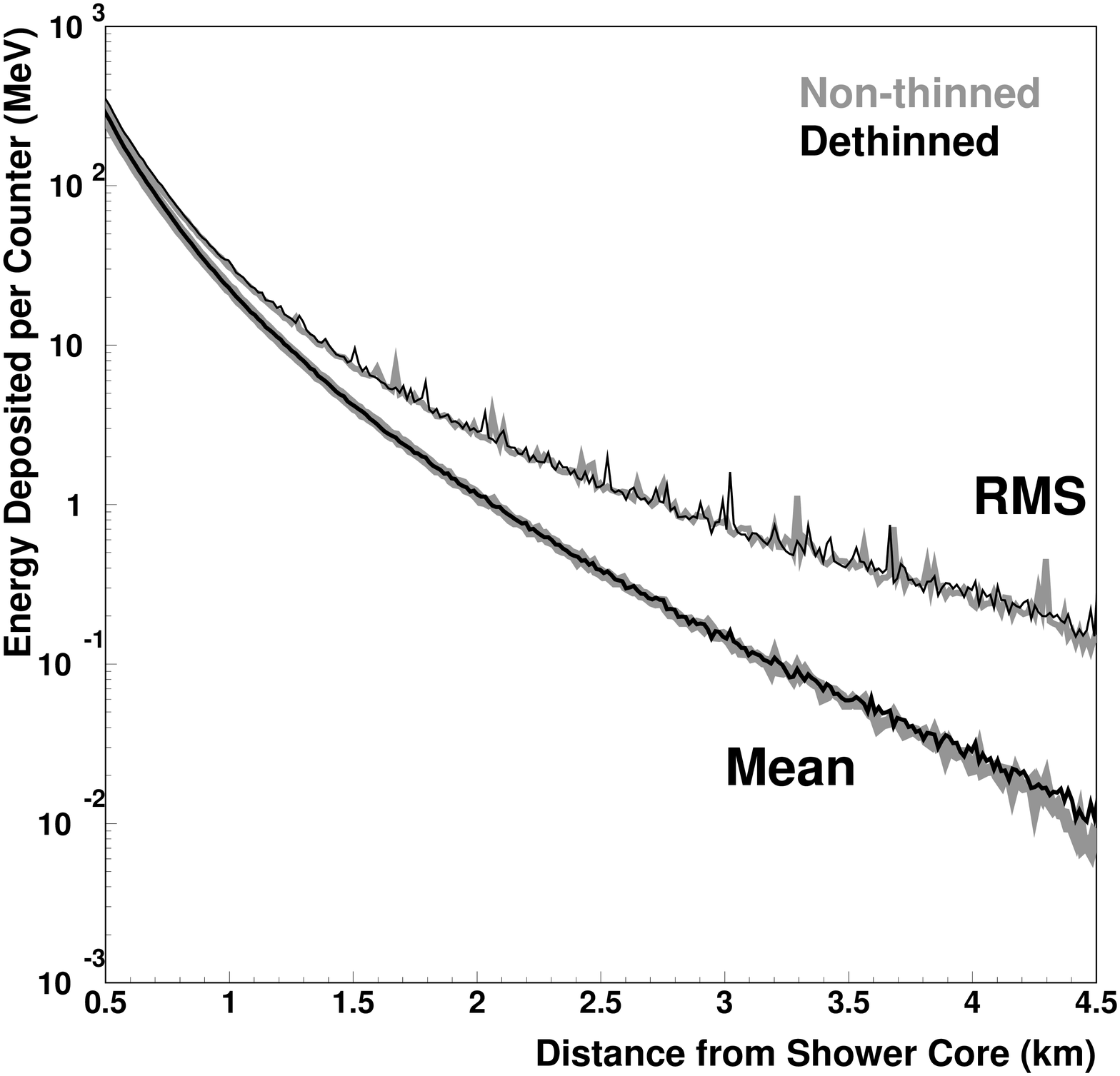}\label{fig:dth_v_nth}}
  \caption{A comparison of energy deposition per counter versus perpendicular distance-to-core for a non-thinned 
  and a thinned simulation before (left) and after (right) the dethinning 
  procedure is applied.  Both simulations are of a proton with a primary 
  energy of $10^{19}$~eV and a primary zenith angle of $45^\circ$.  While the 
  mean energy deposition agrees in all cases, the variation in the energy deposition (RMS) shows
  much better agreement after dethinning.}
  \label{fig:dethinning}
\end{figure*}

Each shower in the CORSIKA library is then
subjected to dethinning~\cite{dethinned}.  For each simulated event, all shower particles that strike the ground are divided spatially by their landing spots into $6\times6{\rm m^2}$ ``tiles'' on the desert floor and into $20{\rm ns}$ wide bins by their arrival time.  The total energy deposited by all particles that landed in a particular tile, and into a virtual TA SD counter located at its center, is calculated using the 
GEANT4 simulation package~\cite{geant4}.  Note this analysis assumes many more virtual SD counters (spaced every 6 m instead of 1.2 km) than are actually present in the experiment.  Back scattering of particles striking the ground within the tile is included in the simulation.  The energy deposited as a 
function of time is stored in the shower library.  Figure~\ref{fig:dethinning} 
shows the comparison of energy deposition in SD counters vs. distance-to-core from a simulated $10^{19}$ eV shower before and after de-thinning.  The plot on the right, made using a de-thinned shower, shows excellent agreement to an identical unthinned shower in both the mean energy deposit and its RMS variation, plotted as functions of distance-to-core.  In contrast, the same plot on the left comparing the same shower after thinning to the same unthinned shower shows a discrepancy in the RMS variation in energy deposition by up to an order of magnitude.

In the concluding step of the shower library generation, each tiled shower
is sampled ~2000 times through a detailed simulation of the detector, including electronics.  The shower core positions, the azimuth of the shower axis, and event times are varied in this process.   The detector simulation utilizes real-time calibration information from the TA
SD to effect a highly detailed, time-specific 
simulation of the detector operating conditions.
Additionally, random background particles are inserted into the electronics
readout based on secondary flux derived from additional CORSIKA simulations
of the low-energy cosmic ray spectrum reported by the BESS 
Collaboration~\cite{bess}. The net 
result of this step is to convert each dethinned CORSIKA shower into an event
library of
simulated detector events in a data format identical to that produced by the 
TA SD instrumentation.

\begin{figure*}[t,b]
  \begin{center}
  \includegraphics[width=\textwidth]{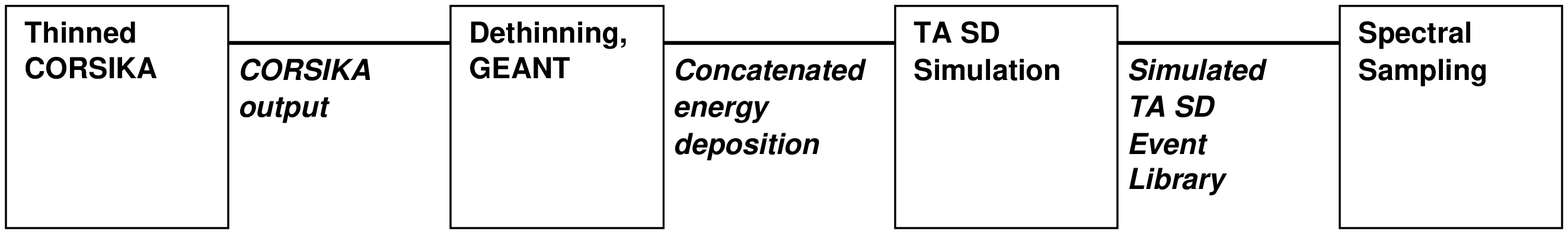}
  \end{center}
  \caption{Steps for simulating the TA SD data set.  Each box represents
  one or more computational routines used to produce the input files required
  for the next step.}
  \label{figure:flowchart}
\end{figure*}

In order to achieve a highly accurate representation of the actual TA SD data set, we sample simulated events from our event library with a primary energy distribution and composition according to published HiRes energy spectrum~\cite{hiresmono:gzk} and composition~\cite{hires_comp}, respectively.
The resulting MC event set is then processed by the same analysis
program as the TA SD data.  This process chain is illustrated by the diagram in
Figure~\ref{figure:flowchart}.

Finally, we validate the accuracy of the simulation by comparing distributions
of key observables obtained from the MC events with those from real data.  As will be seen in the next section, our de-thinned shower samples give excellent agreement  in these data-MC comparisons, thereby verifying the reliability of detector resolutions and of the detector acceptance calculation obtained from our simulation algorithm. The primary advantage of this process lies in that the trigger efficiency, reconstruction quality cuts, and the effects of finite resolution~\cite{unfolding}\cite{hiresmono:dtmc} are all automatically included in the analysis.

\begin{figure}[t,b]
  \centering
  \includegraphics[width=0.5\textwidth]{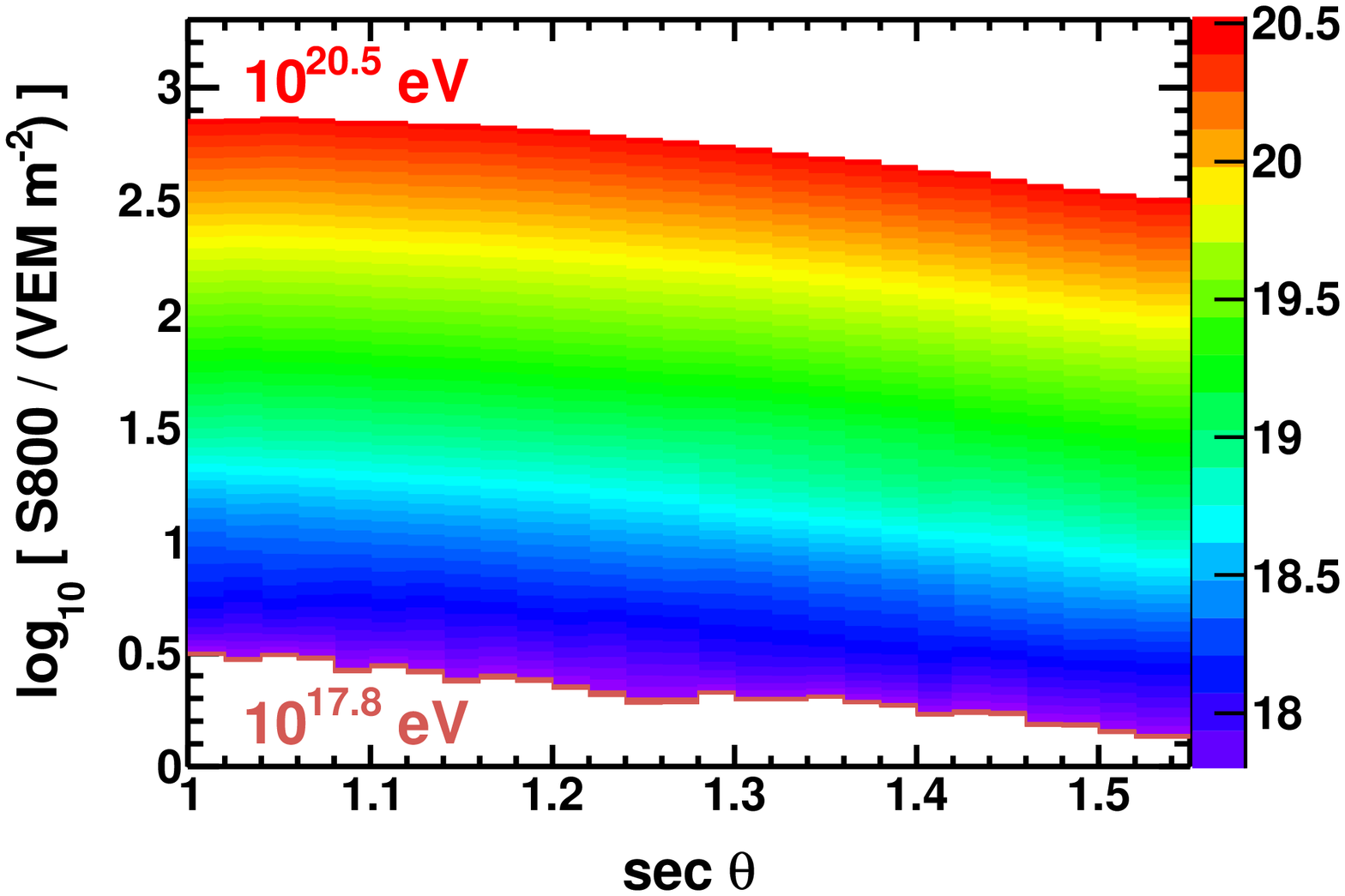}
  \caption { Energy as a function of reconstructed $S800$ and
    $\sec\theta$ made from the dethinned MC event set.  The true (input) values of the primary energy are represented (along the z-axis) by color according to the key shown to the right.}
  \label{figure:entable}
\end{figure}

A functional relationship between $S800$, the primary zenith angle ($\theta$), and the primary energy is constructed using the de-thinned Monte Carlo Event set. Each simulated event is subjected to the same geometrical reconstruction as described above, and the value of $S800$ obtained in the same way. A three-dimensional scatter plot is then made of the input (generated) primary energy of each shower plotted in the $z$-direction, vs. $\sec\theta$ in the $x$-direction, and the logarithm of the $S800$ value in the $y$-direction. The points in this plot form a surface that represents the shower energy as a bi-variate function of $\sec\theta$ and $\log_{10}(S800)$.  The function obtained for this work is shown in
Figure~\ref{figure:entable}, 
in which the value of energy is represented by color according to the key attached to the right of the plot. The information contained in Figure~\ref{figure:entable} is used to determine the energy of both real and simulated events from the interpolated $S800$ values.  

\section{Data - Monte Carlo Comparisons}

A crucial part of the TA SD simulation program is the comparison of data and MC distributions. The success of these comparisons validates the accuracy of the simulated event set in its representation of the real data set, and demonstrates the reliability of analysis procedures that depend on the Monte Carlo. These include the construction of energy vs. $\sec\theta$ and $S800$ plot in Figure~\ref{figure:entable}, the determination of detector resolutions, and the acceptance calculation.

\begin{figure*}[t,b]
  \centering
  \subfloat[Data]{\includegraphics[width=0.5\textwidth]{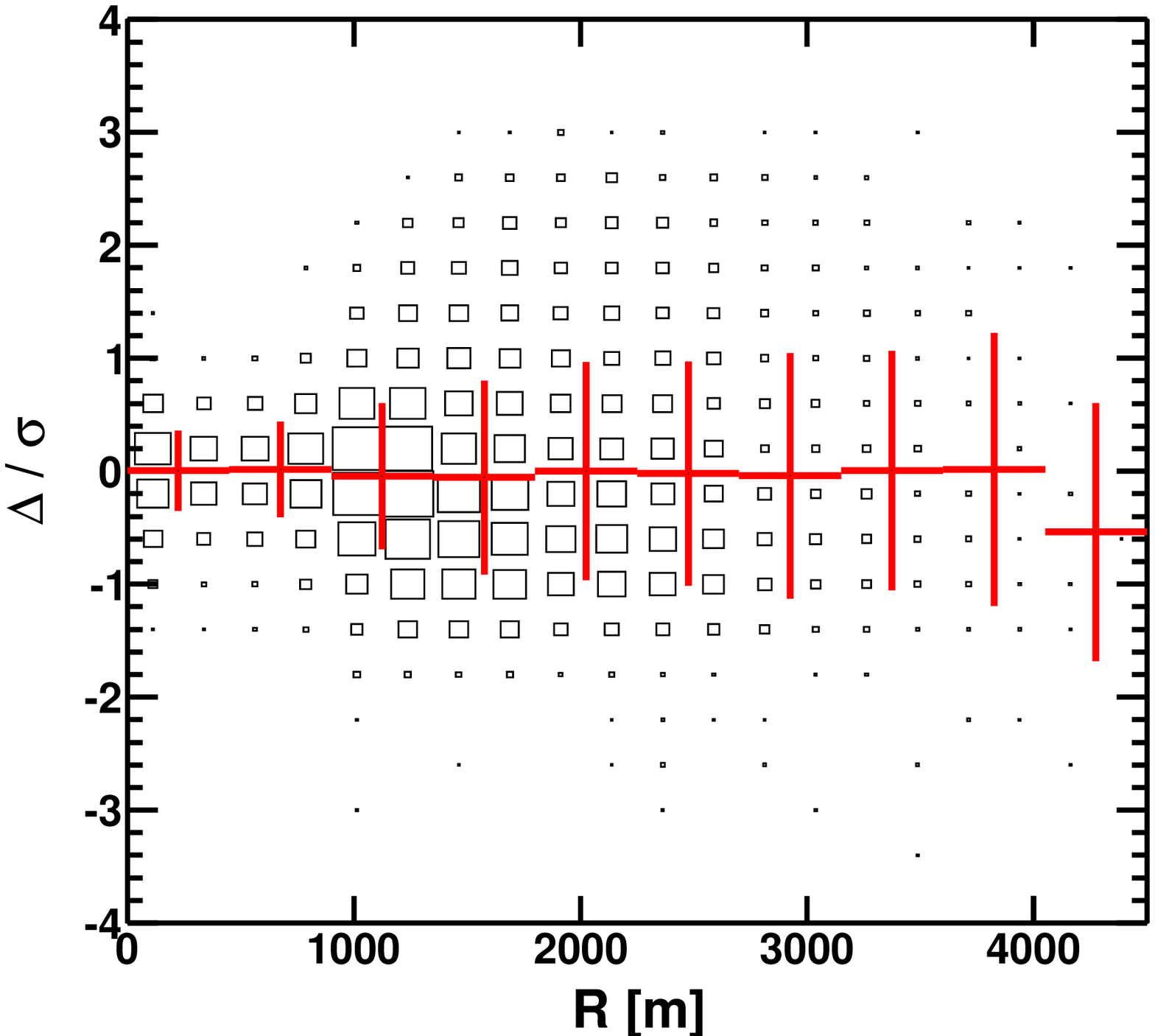}\label{figure:dt_tfit_rsd}}
  \subfloat[MC]{\includegraphics[width=0.5\textwidth]{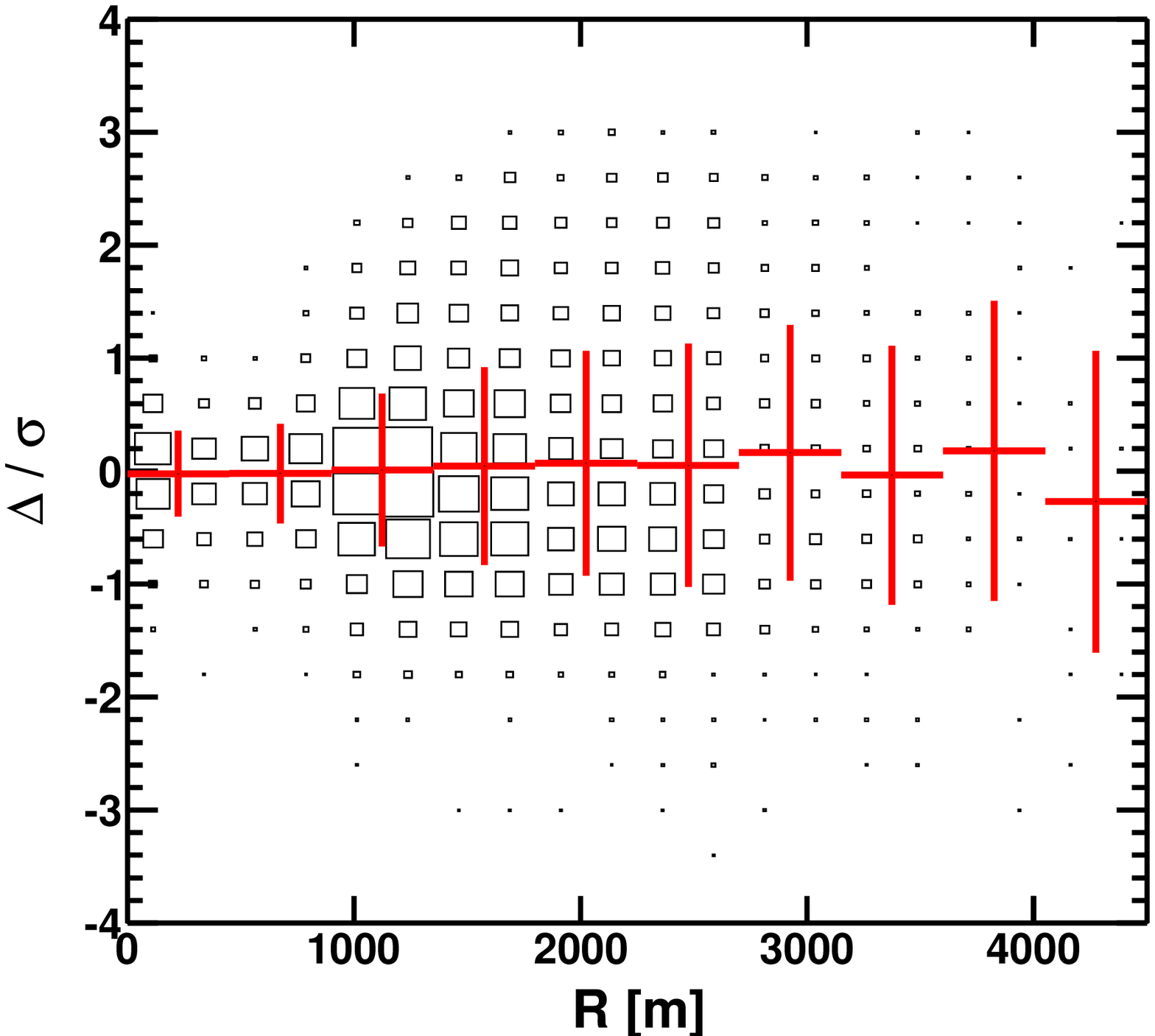}\label{figure:mc_tfit_rsd}}
  \caption{ Time fit residual normalized by the uncertainty plotted
    versus the lateral distance from the shower core. Each entry is a
    counter that is a part of the event and the plot is made over all
    events in the data and MC sets.  The superimposed points with error
    bars are the profile showing the mean
    and the RMS of the normalized residual. }
  \label{figure:tfit_rsd}
\end{figure*}

\begin{figure*}[t,b] 
  \centering 
  \subfloat[Zenith Angle]{\includegraphics[width=0.3\textwidth]{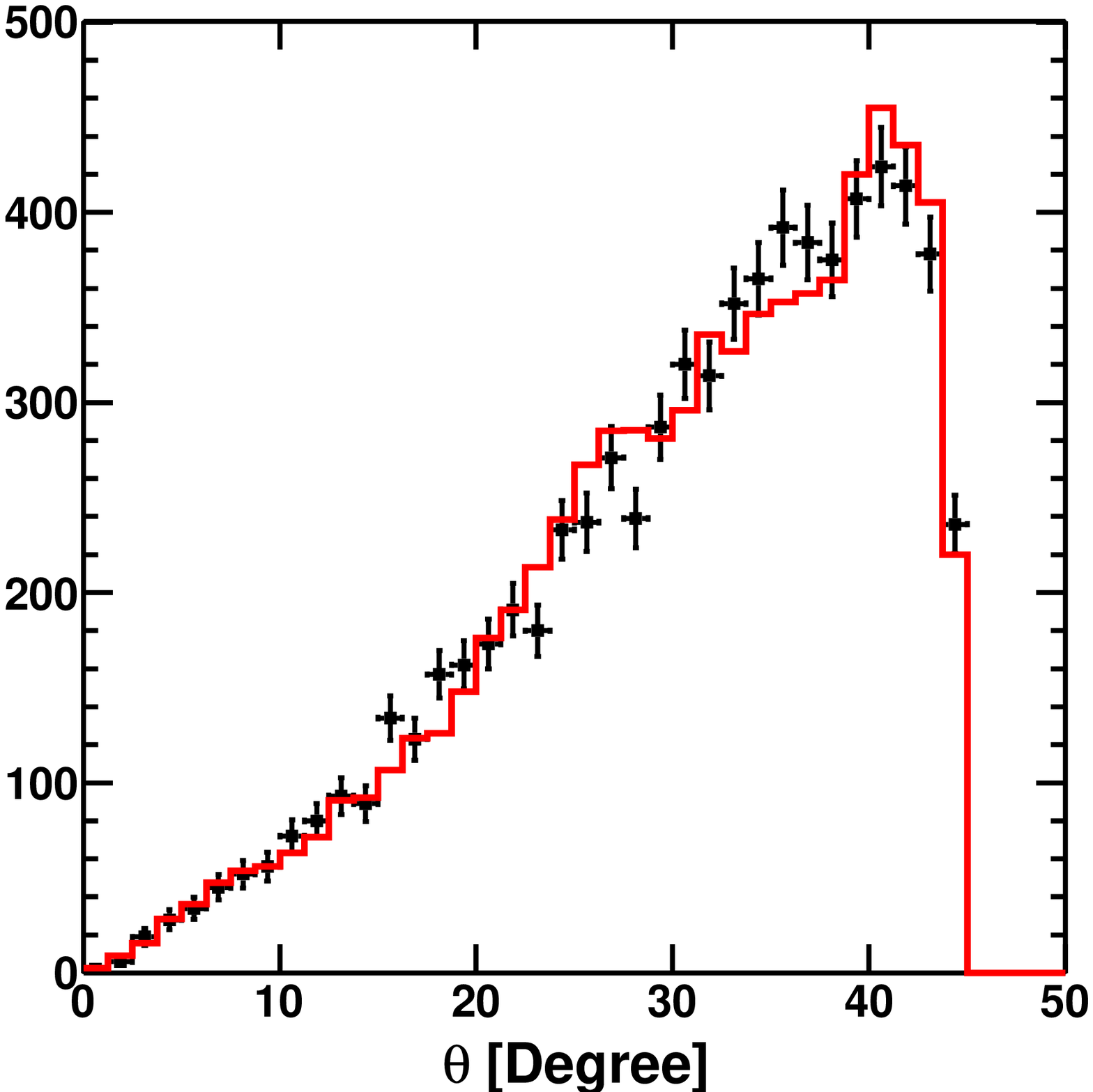}\label{figure:dtmc_theta}}
  \subfloat[Azimuthal Angle]{\includegraphics[width=0.3\textwidth]{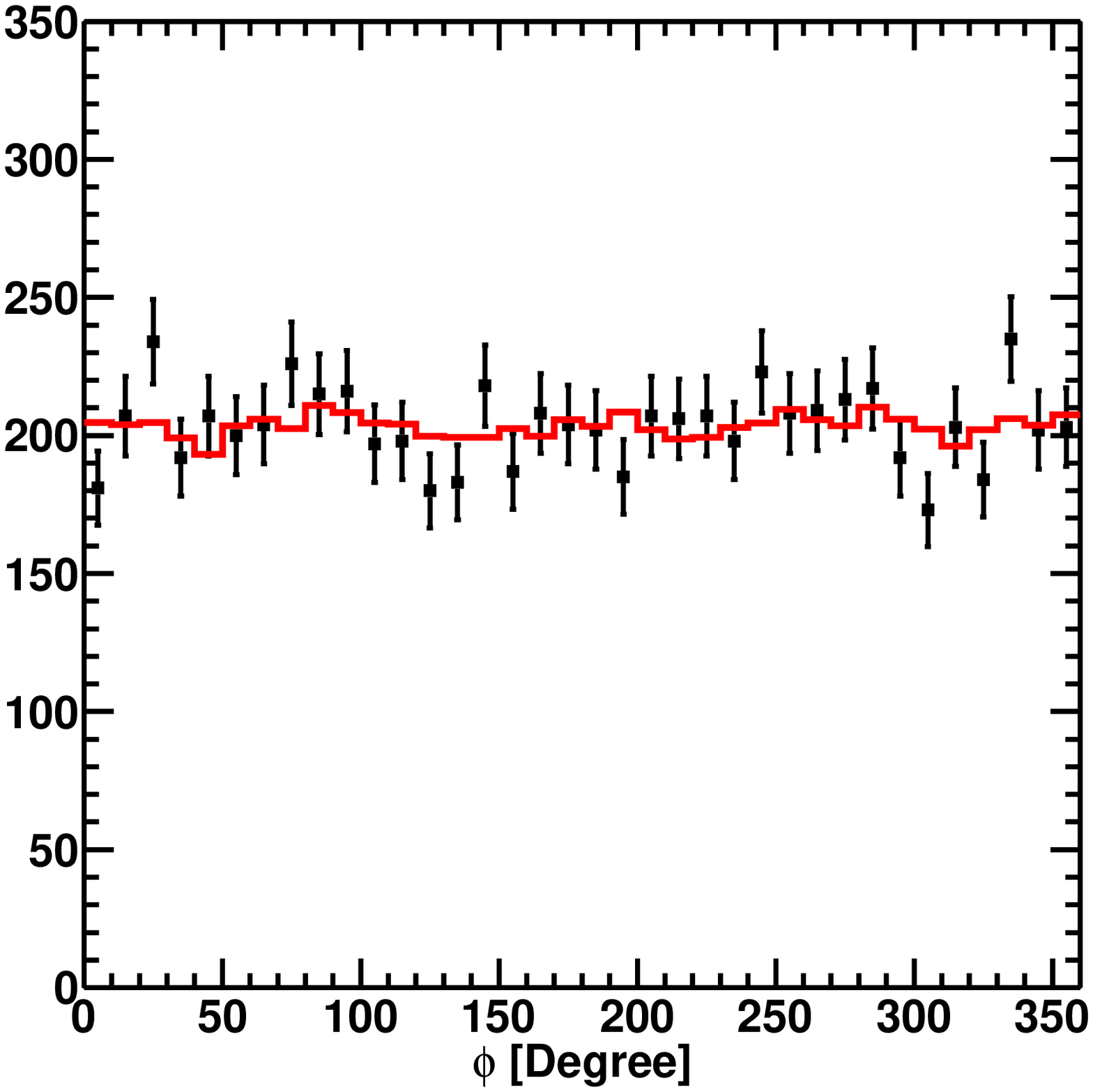}\label{figure:dtmc_phi}}
  \subfloat[Core X]{\includegraphics[width=0.3\textwidth]{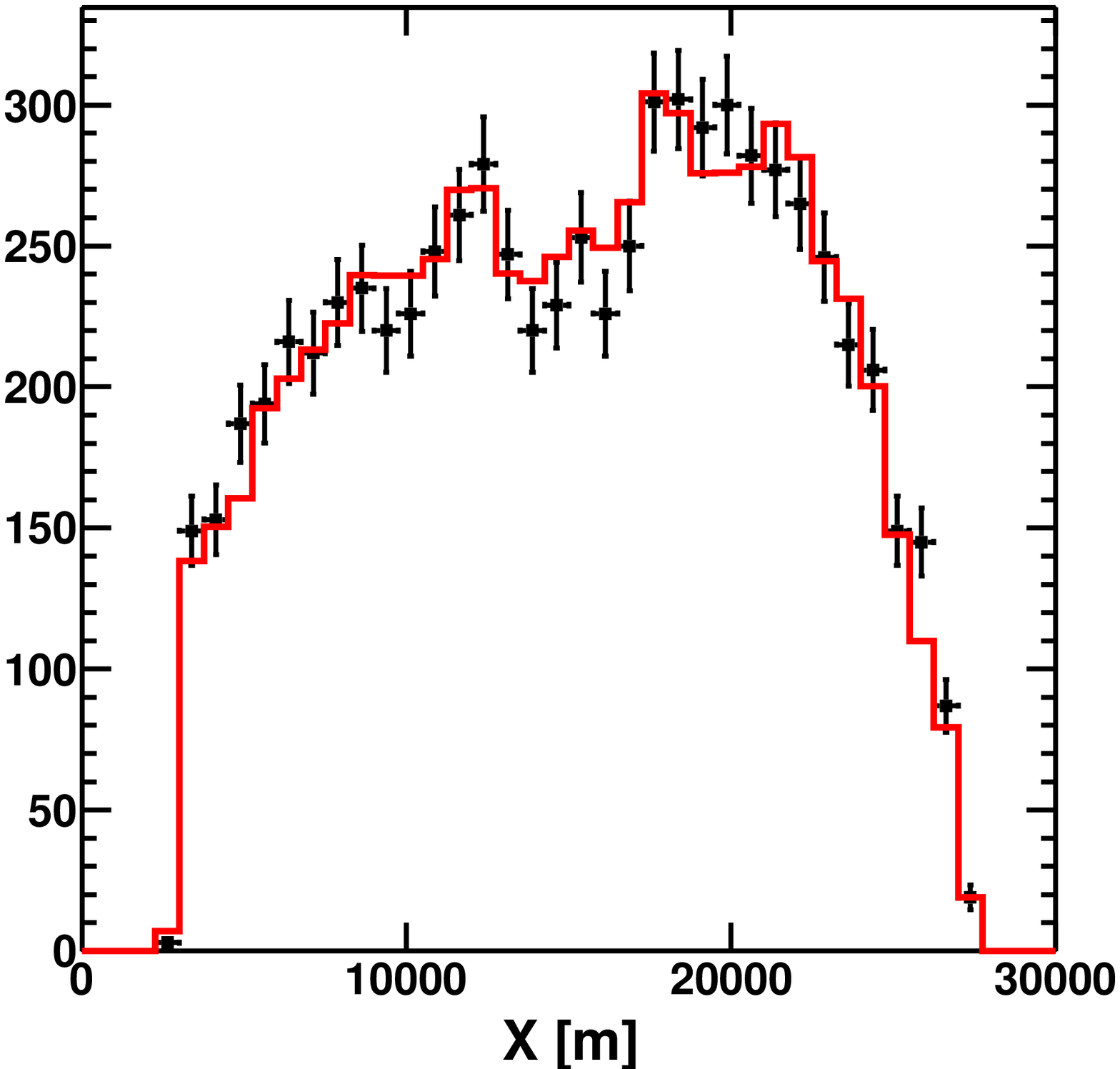}\label{figure:dtmc_xcore}}
  \linebreak
  \subfloat[Core Y]{\includegraphics[width=0.3\textwidth]{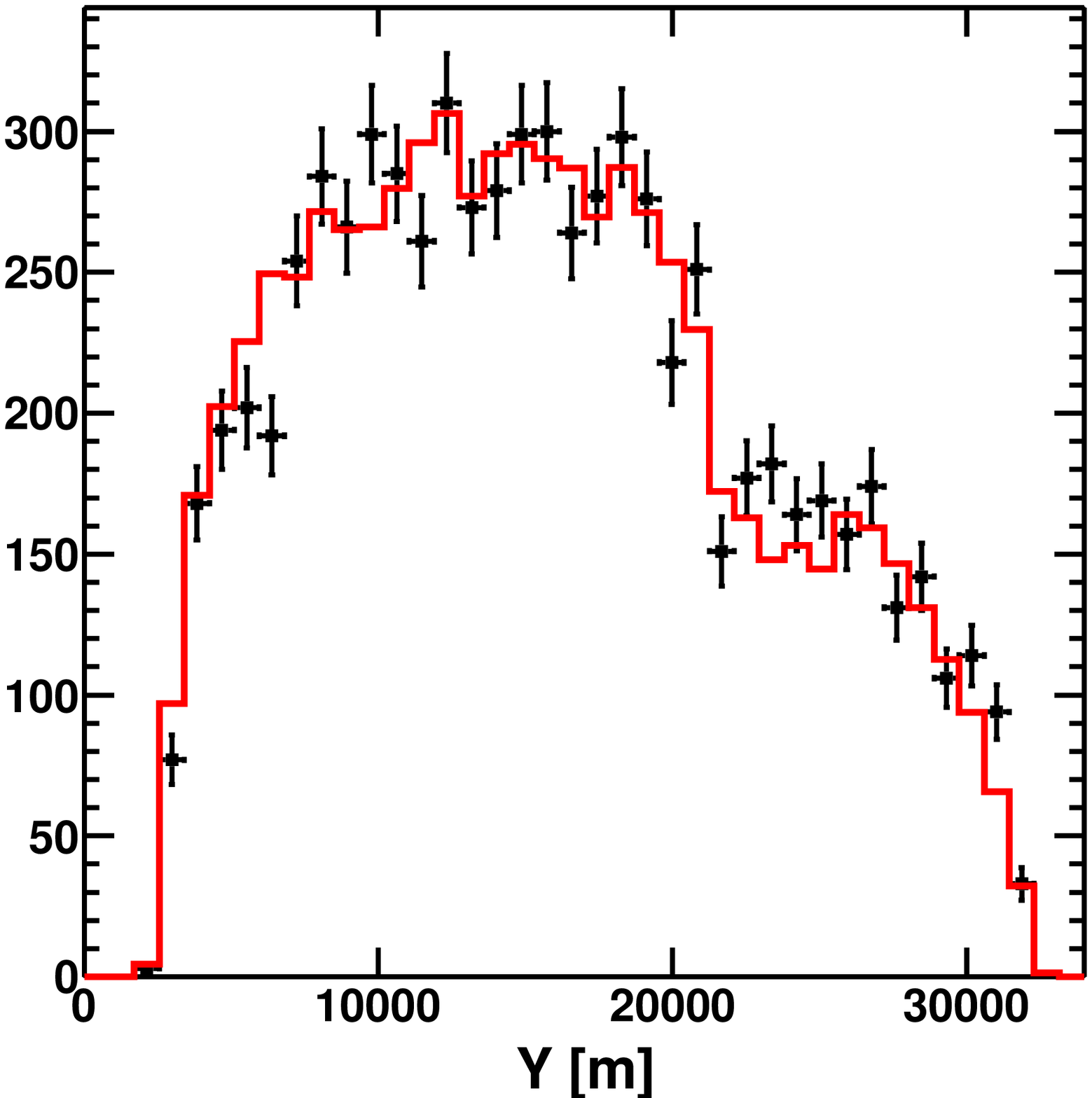}\label{figure:dtmc_ycore}}
  \subfloat[Right Ascension]{\includegraphics[width=0.3\textwidth]{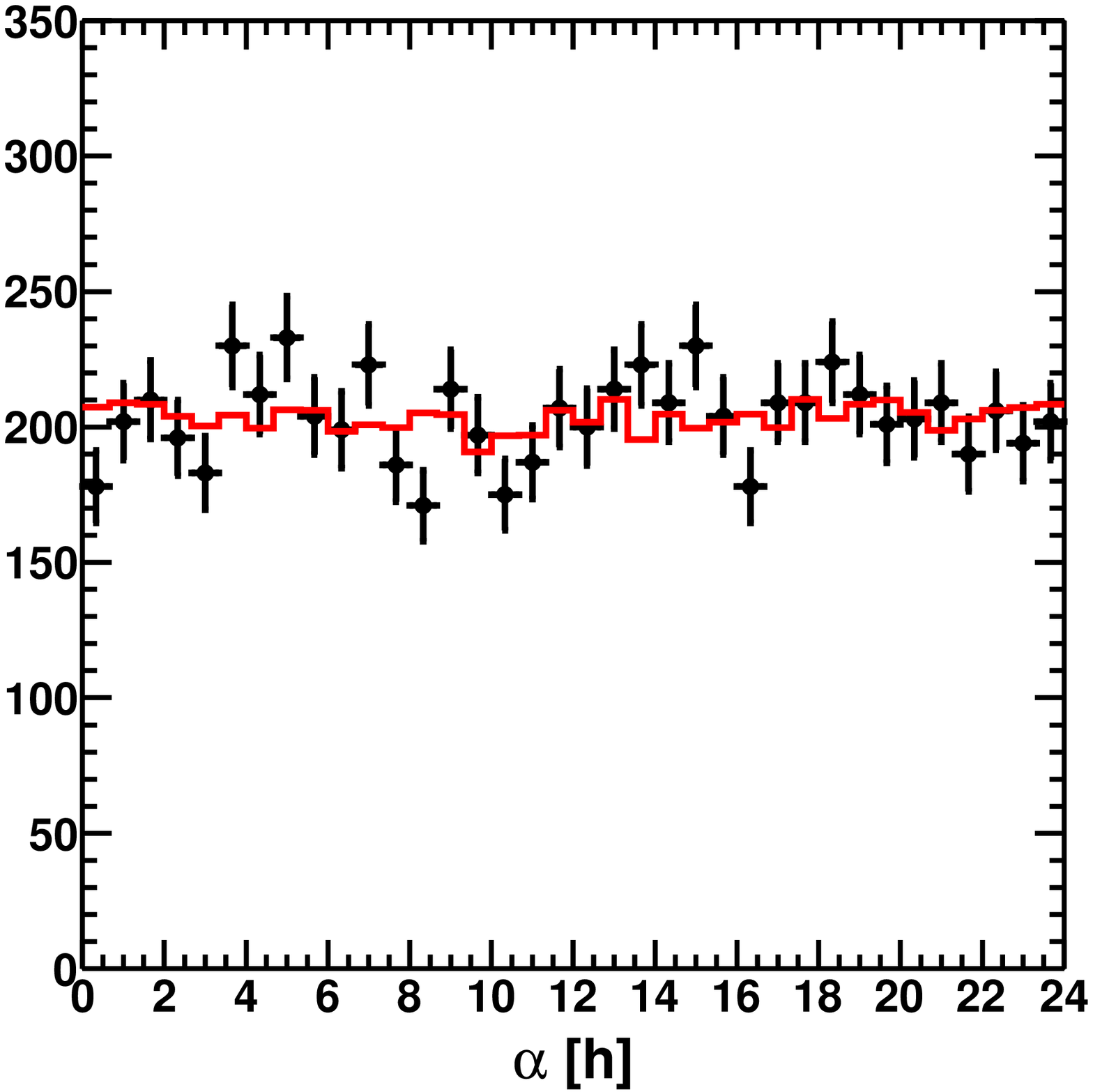}\label{figure:dtmc_alpha}}
  \subfloat[Declination]{\includegraphics[width=0.3\textwidth]{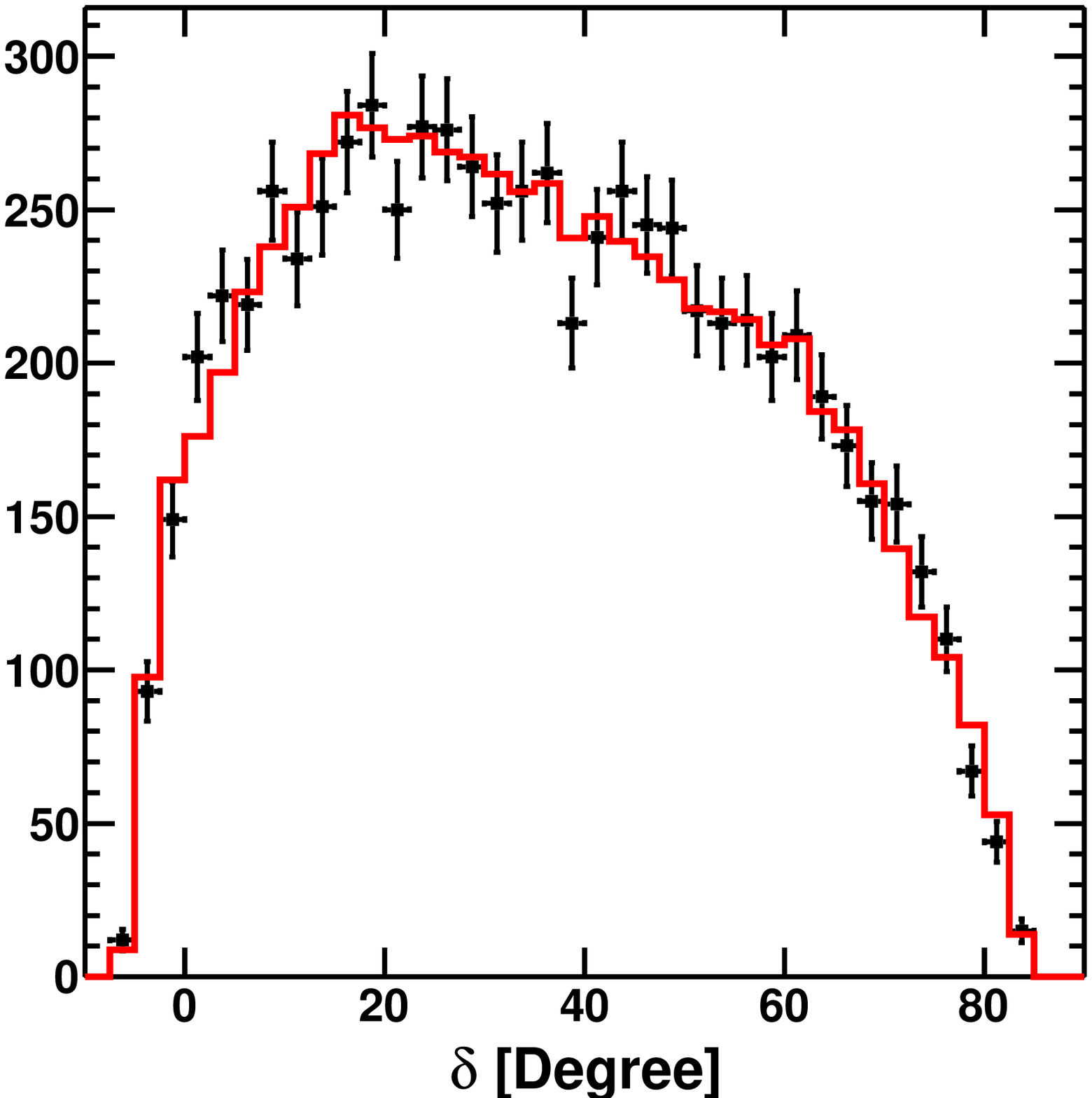}\label{figure:dtmc_delta}}
  \caption{ Data and MC comparison of the geometrical quantities.
    Points with error bars are the data and superimposed solid lines are
    the MC histograms normalized to the same integral as the data. }
  \label{figure:dtmc_geom}
\end{figure*}

Figure~\ref{figure:tfit_rsd} 
shows normalized residuals of the time
fit as a function of lateral distance from the shower core 
(i.e. the difference between the SD counter time and the fit
value, all divided by the uncertainty in counter time).   
In Figure~\ref{figure:mc_tfit_rsd}, we apply the same time delay function to 
simulated events.  In Figure~\ref{figure:dtmc_geom}, we compare the real and 
simulated distributions for several geometric observables.
By considering the comparison of 
Figures~\ref{figure:tfit_rsd}~and~\ref{figure:mc_tfit_rsd} in conjunction with 
the further comparisons shown in Figure~\ref{figure:dtmc_geom}, 
we establish
that the simulated event set possesses a distribution of geometric 
characteristics that are very similar to those of the real data.

\begin{figure*}[t,b]
  \centering 
  \subfloat[Signal per counter]{\includegraphics[width=0.3\textwidth]{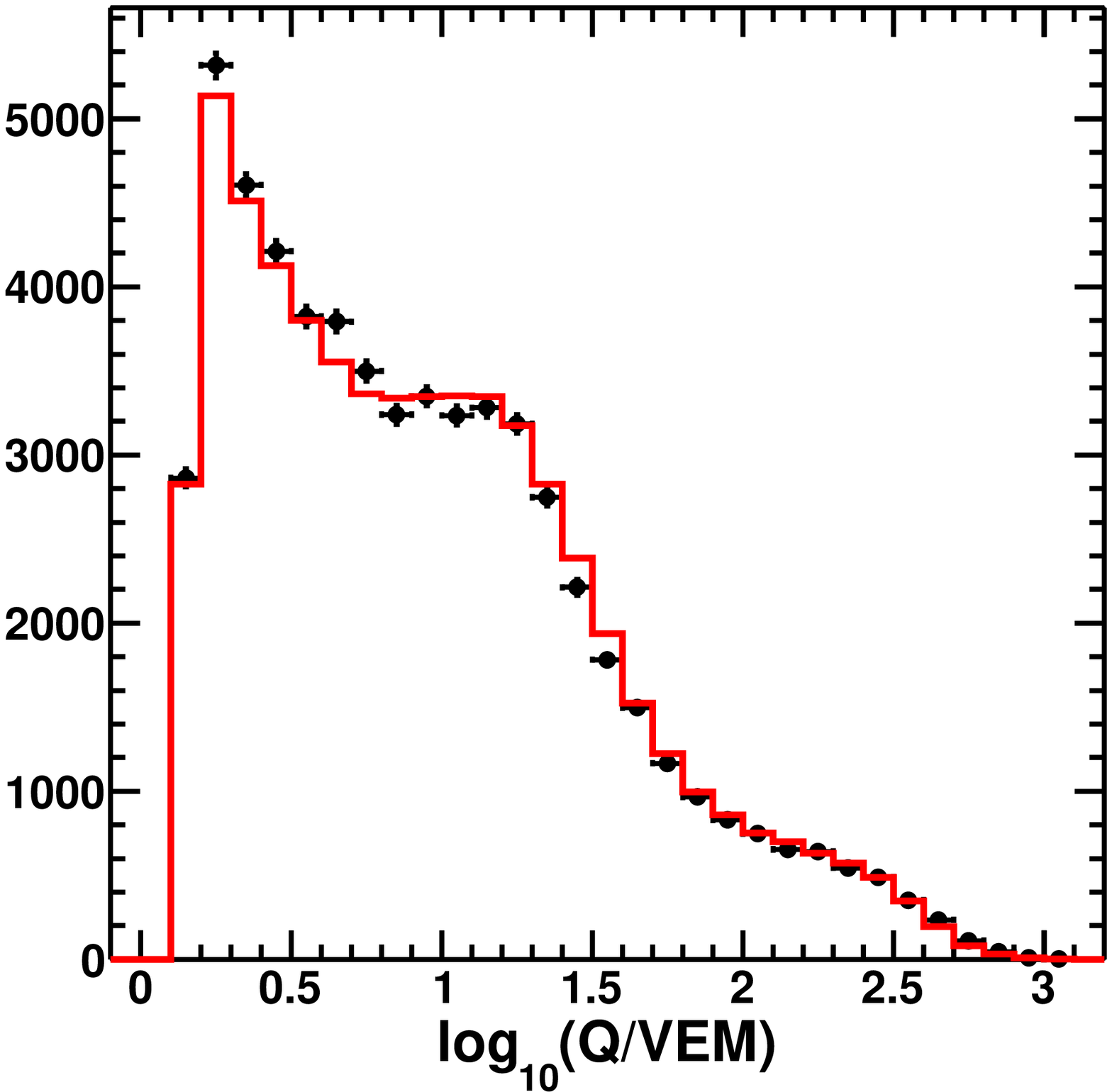}\label{figure:dtmc_vempcnt}}
  \subfloat[Signal per event]{\includegraphics[width=0.3\textwidth]{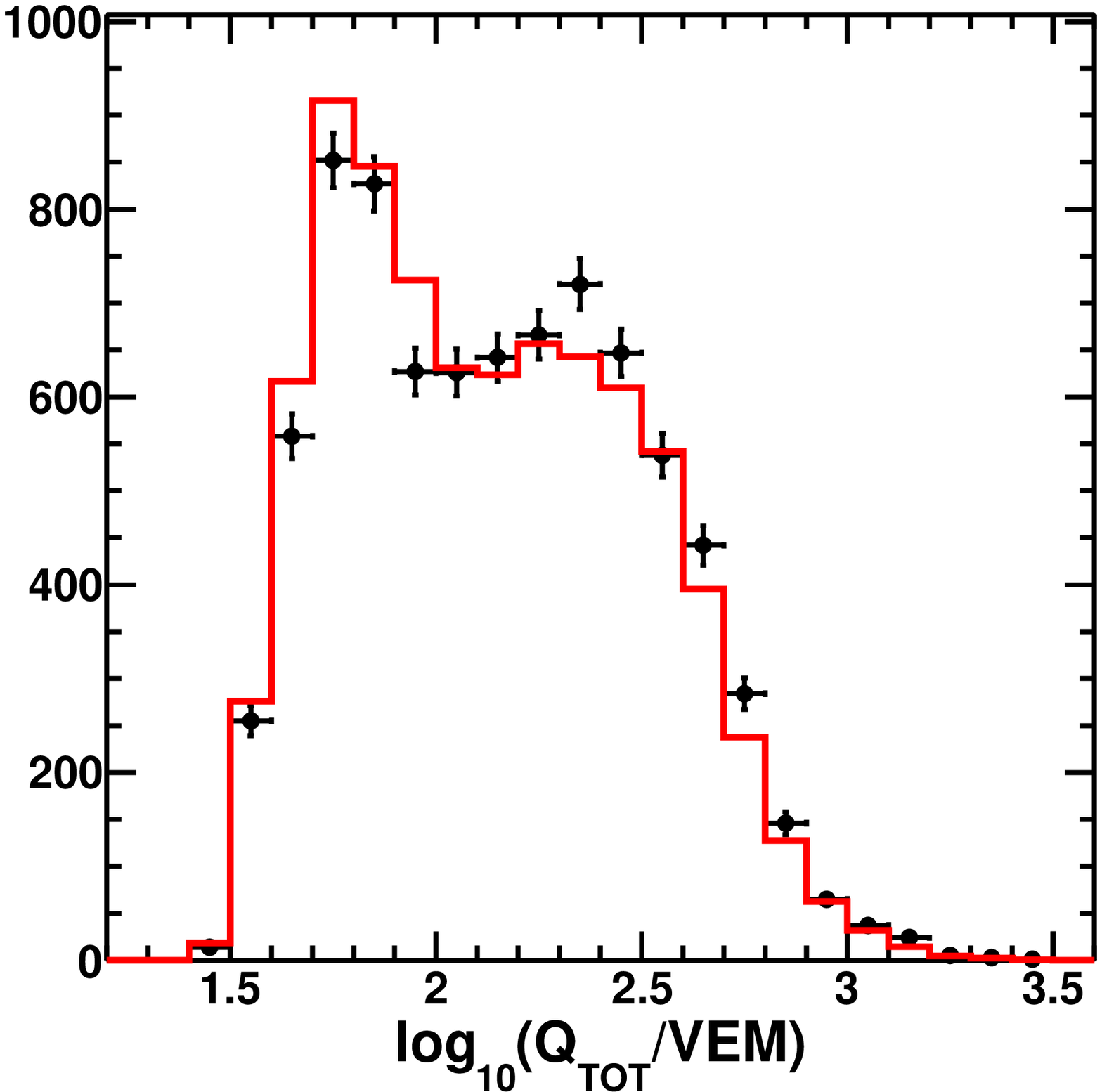}\label{figure:dtmc_vempevt}}
  \subfloat[LDF $\chi^{2}/dof$]{\includegraphics[width=0.3\textwidth] {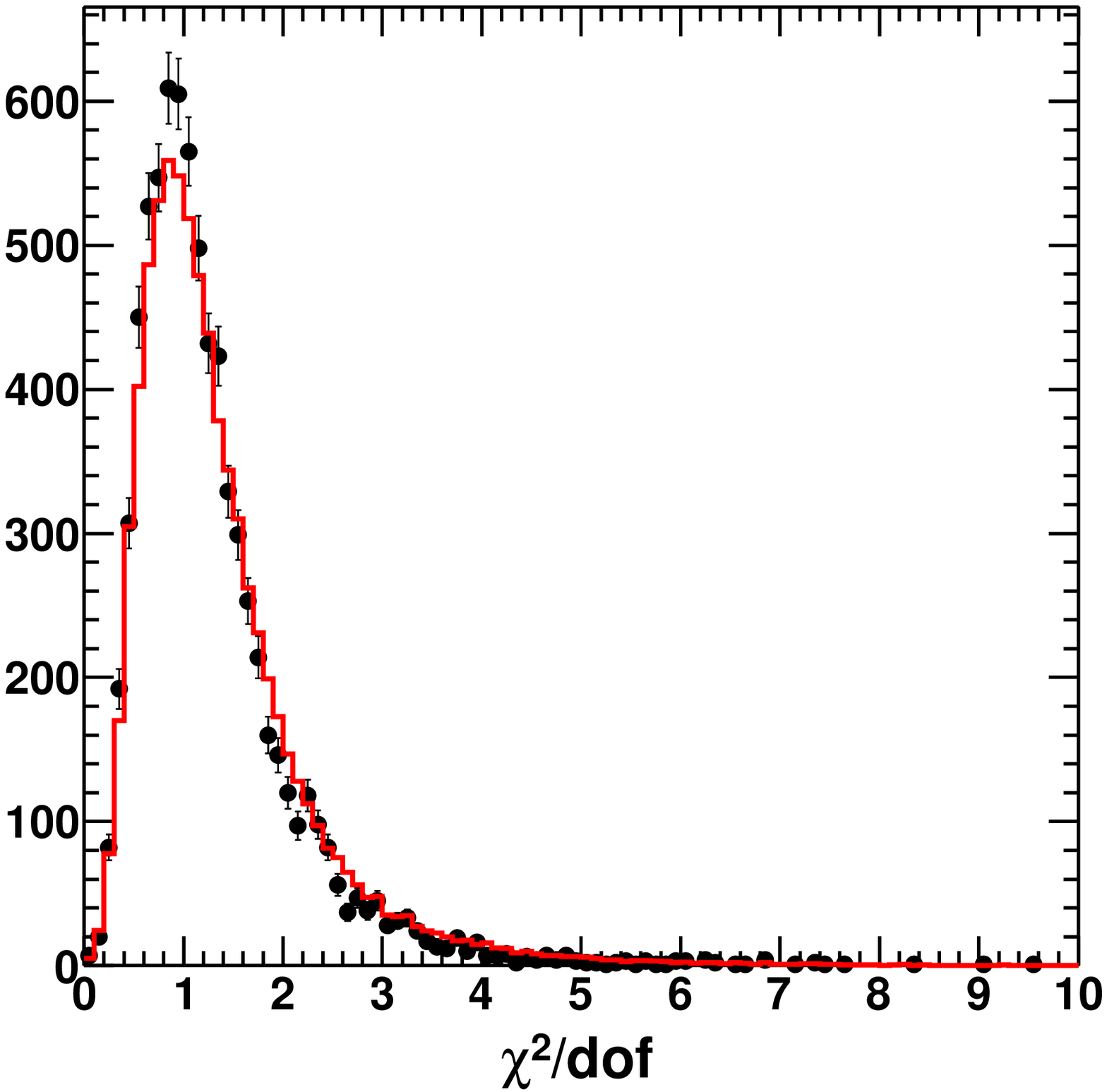}\label{figure:dtmc_ldfchi2pdof}}
  \linebreak
  \subfloat[S800]{\includegraphics[width=0.3\textwidth]{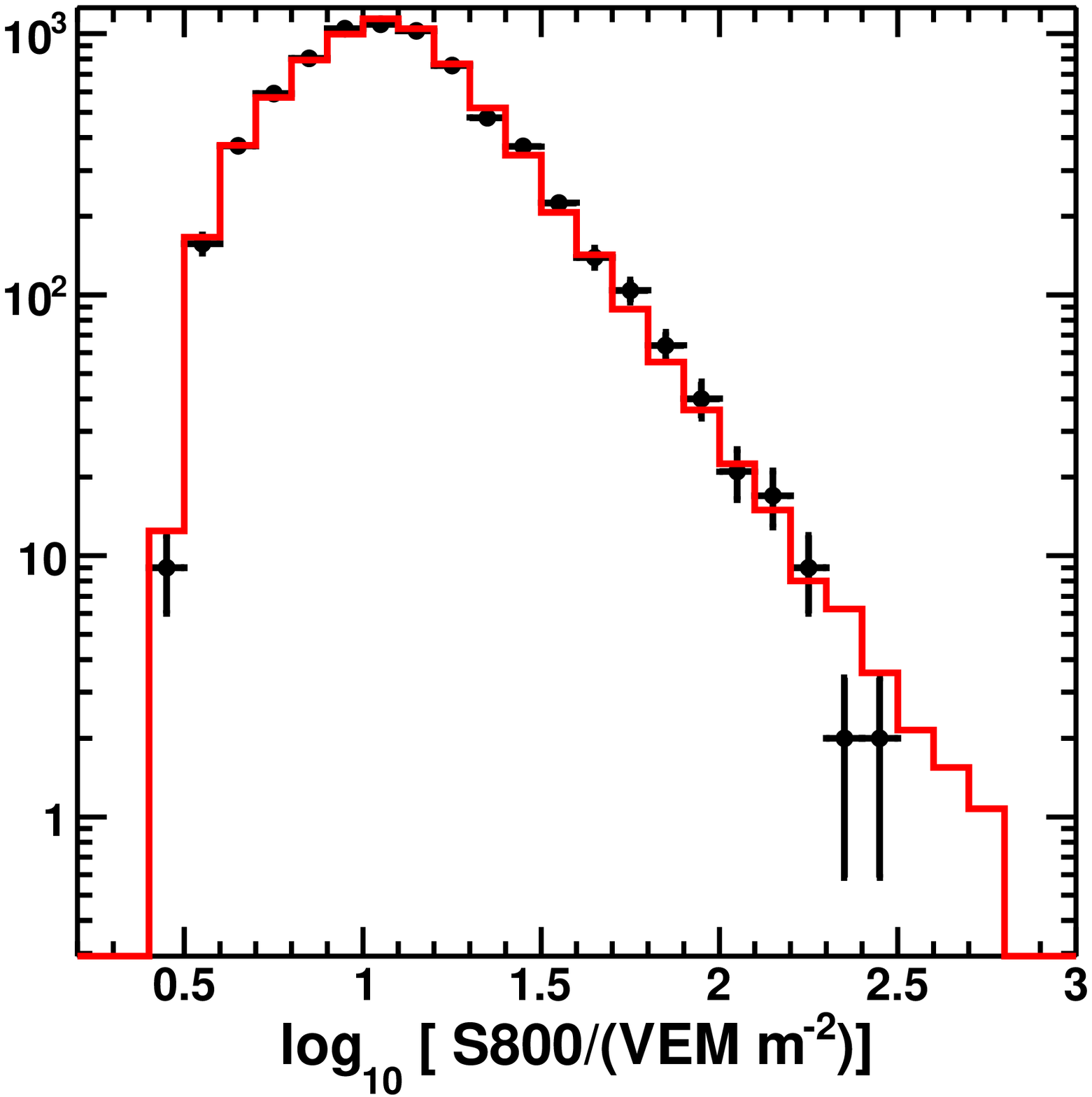}\label{figure:dtmc_s800}}
  \subfloat[Energy]{\includegraphics[width=0.3\textwidth]{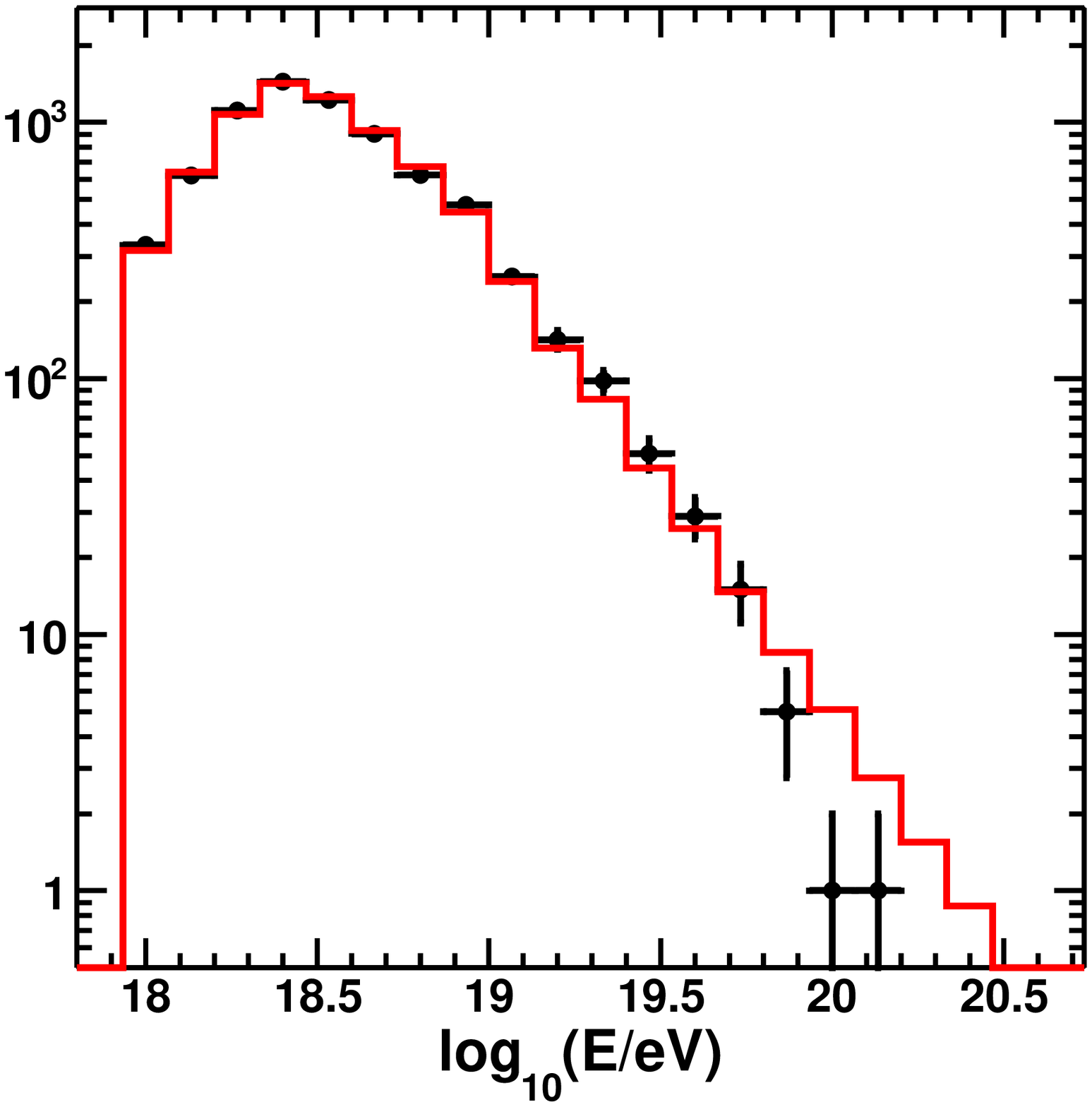}\label{figure:dtmc_energy}}
  \linebreak
  \caption{ Data and MC comparison of the quantities related to the lateral distribution. Points with error bars are the data and superimposed solid lines are
    the MC histograms normalized to the same integral as the data. }
  \label{figure:dtmc_ldf}
\end{figure*}

Figure~\ref{figure:dtmc_ldf} 
shows the data-MC comparison of
real and simulated lateral distribution quantities and the reconstructed energy. It should be noted that while the comparison 
Figures~\ref{figure:dtmc_s800}~and~\ref{figure:dtmc_energy} show a small deficit of events at
large values of $S800$ and energy in the data, this difference is expected because we do not include the simulation of the GZK suppression~\cite{greisen}\cite{zk} at this level of analysis, and it does not reflect a fundamental disagreement between the real and simulated event sets.  The data-Monte Carlo comparison plots show excellent agreement.

\section{Detector Resolutions}

The detector resolutions are determined by comparing the reconstructed and generated values for those simulated showers that survived the event selection cuts described in the previous section. The two key resolutions of interest are those of the arrival direction (of particular importance in anisotropy studies) and primary energy (important for the energy spectrum and for anisotropy).

\begin{figure*}[t,b]
  \centerline{
    \subfloat[]{\includegraphics[width=0.33\textwidth]{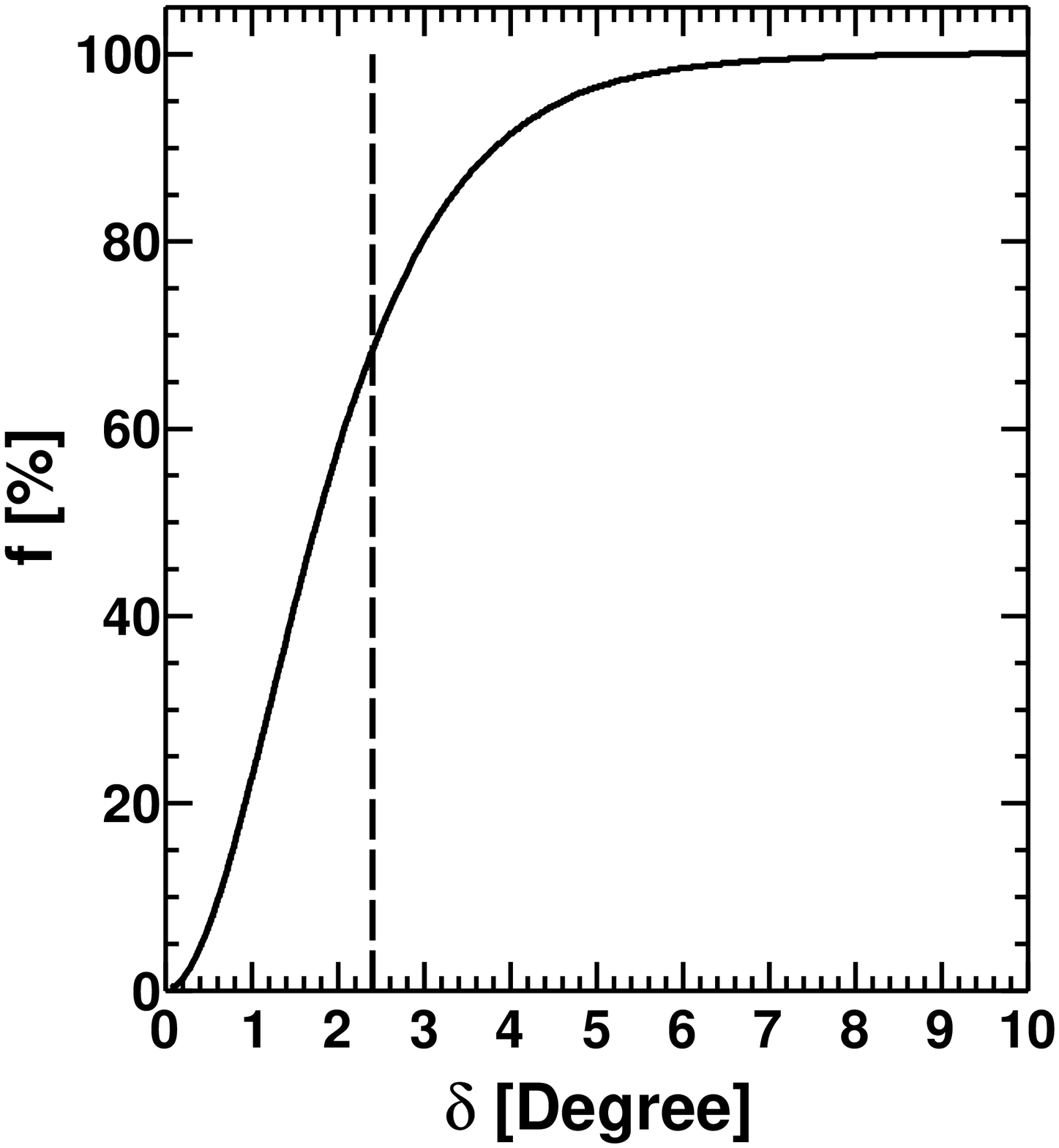}\label{figure:tasd_angres1}}
    \subfloat[]{\includegraphics[width=0.33\textwidth]{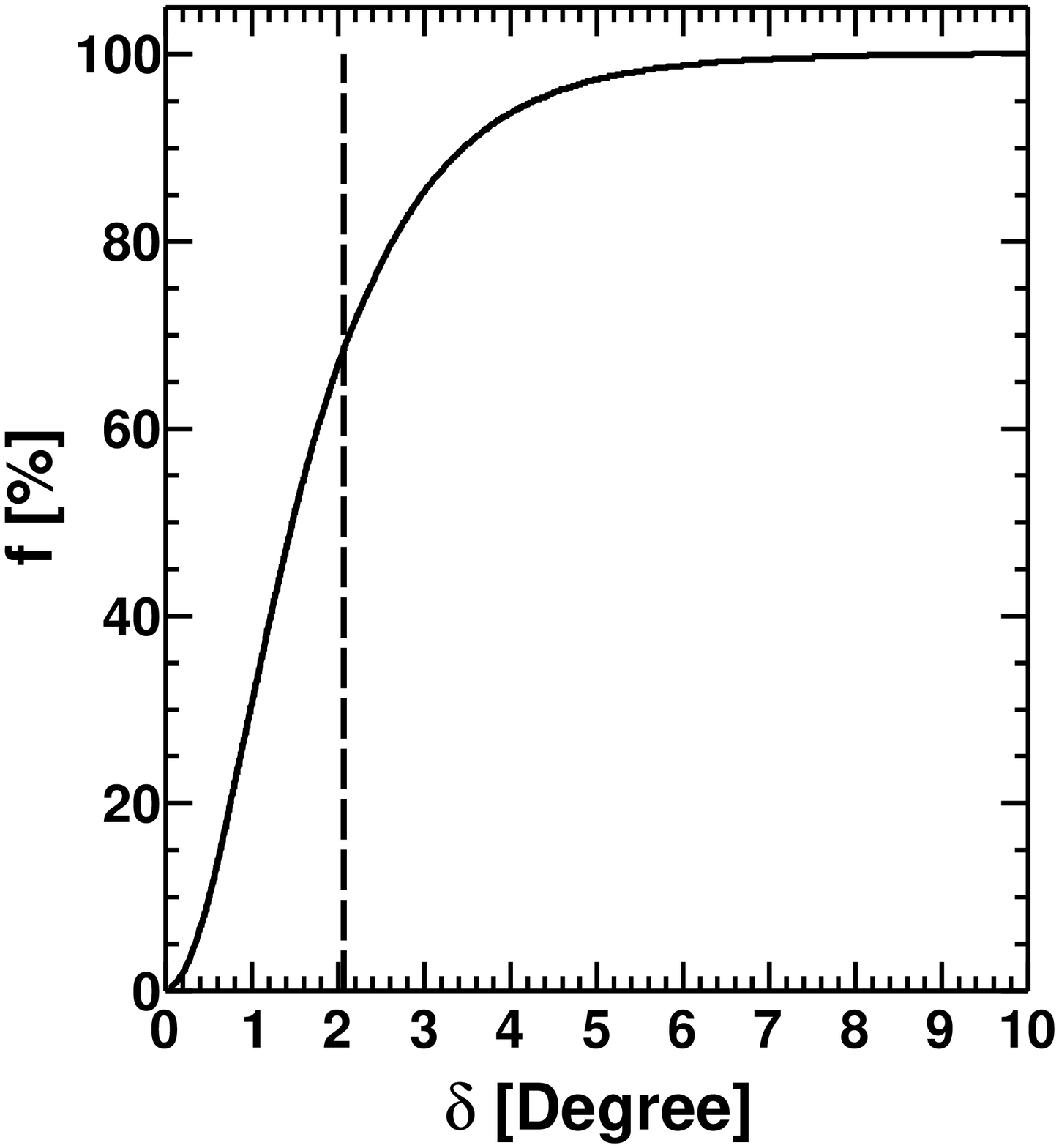}\label{figure:tasd_angres2}}
    \subfloat[]{\includegraphics[width=0.33\textwidth]{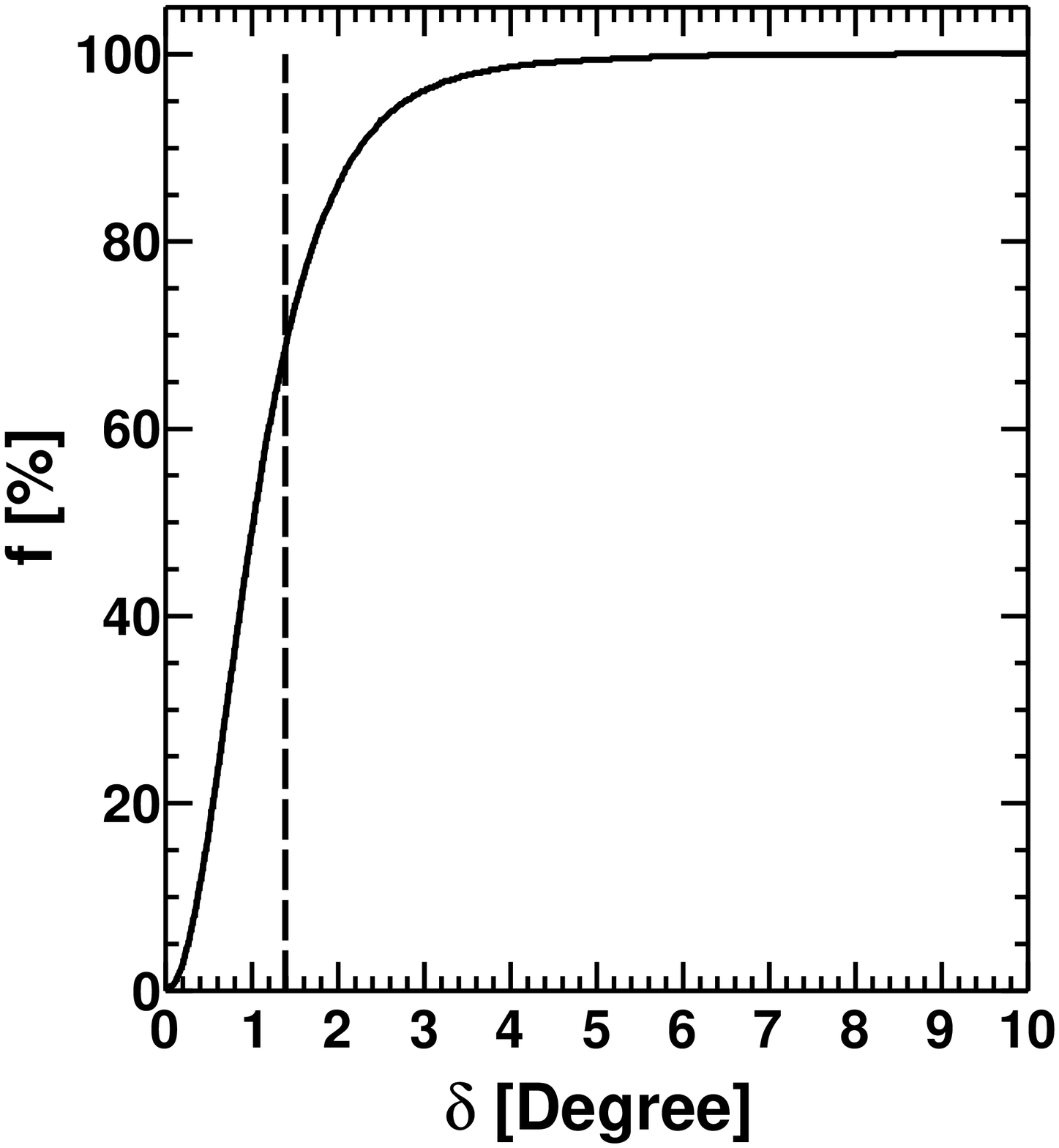}\label{figure:tasd_angres3}}
  }
  \caption{ The TA SD angular resolution evaluated using a Monte-Carlo
    spectral set. Cumulative histograms of the opening angle between
    the reconstructed and the true (MC generated) event directions are
    shown using three energy slices:
    \protect\subref{figure:tasd_angres1} $10^{18.0}\mathrm{eV} < E <
    10^{18.5}\mathrm{eV}$, \protect\subref{figure:tasd_angres2}
    $10^{18.5}\mathrm{eV} < E < 10^{19.0}\mathrm{eV}$,
    \protect\subref{figure:tasd_angres3} $E > 10^{19.0}\mathrm{eV}$.
    The X axis represents the opening angle $\delta$ and the Y axis
    represents the fraction $f$ of events (in percent), reconstructing
    within a given opening angle with respect to their true
    directions.  Dashed lines represents the 68\% confidence limits,
    which are the values of $\delta$ containing 68\% of all
    reconstructed events in each energy range. }
  \label{figure:tasd_angres}
\end{figure*}

The angular resolution is obtained from a cumulative
histogram of the opening angle between the reconstructed event
direction $\mathbf{\hat{n}}_{\mathrm{REC}}$ and the true (MC
generated) direction $\mathbf{\hat{n}}_{\mathrm{GEN}}$:
\begin{equation}
  \label{equation:eOpAng} \delta =
\mathrm{cos}^{-1}(\mathbf{\hat{n}}_{\mathrm{REC}} \, \cdot \,
\mathbf{\hat{n}}_{\mathrm{GEN}})
\end{equation} The unit vectors $\mathbf{\hat{n}}_{\mathrm{REC}}$ and
$\mathbf{\hat{n}}_{\mathrm{GEN}}$ are calculated from the shower zenith
and azimuthal angles (both reconstructed and generated).
Figure~\ref{figure:tasd_angres} shows the cumulative distribution of $\delta$ from a spectral MC set (i.e., one generated according to the published HiRes energy spectrum and composition).  The results are displayed for three energy ranges.  Choosing the 68\% confidence interval for stating the answers,
the TA SD angular resolution values are: 2.4$^{o}$ for
$10^{18.0}\mathrm{eV}<E<10^{18.5}\mathrm{eV}$, 2.1$^{o}$ for
$10^{18.5}\mathrm{eV}<E<10^{19.0}\mathrm{eV}$, and 1.4$^{o}$ for
$E>10^{19.0}\mathrm{eV}$.
For the energy resolution we state the root-mean-square (RMS) deviation for the distribution of $R=E_{\mathrm{REC}}/E_{\mathrm{GEN}}$, the ratio of 
the reconstructed ($E_{\mathrm{REC}}$) to the generated
($E_{\mathrm{GEN}}$) event energies.  However, for the display of the distribution of $R$, and for calculating the RMS resolution, it is advantageous to
histogram the natural logarithm of the energy ratio, $R$, because $\ln R$ treats fractional
under-reconstruction and over-reconstruction of event energies in a symmetric
way.  In contrast, a histogram of just the ratios, $R$, would not properly account
for those events with under-reconstructed energies, because $R$ is artificially bounded at zero on the low side, but unbounded on the high side.  This bias can lead to an understated RMS value and hence overstated resolution.

\begin{figure*}[t,b]
  \centerline{
    \subfloat[]{\includegraphics[width=0.33\textwidth]{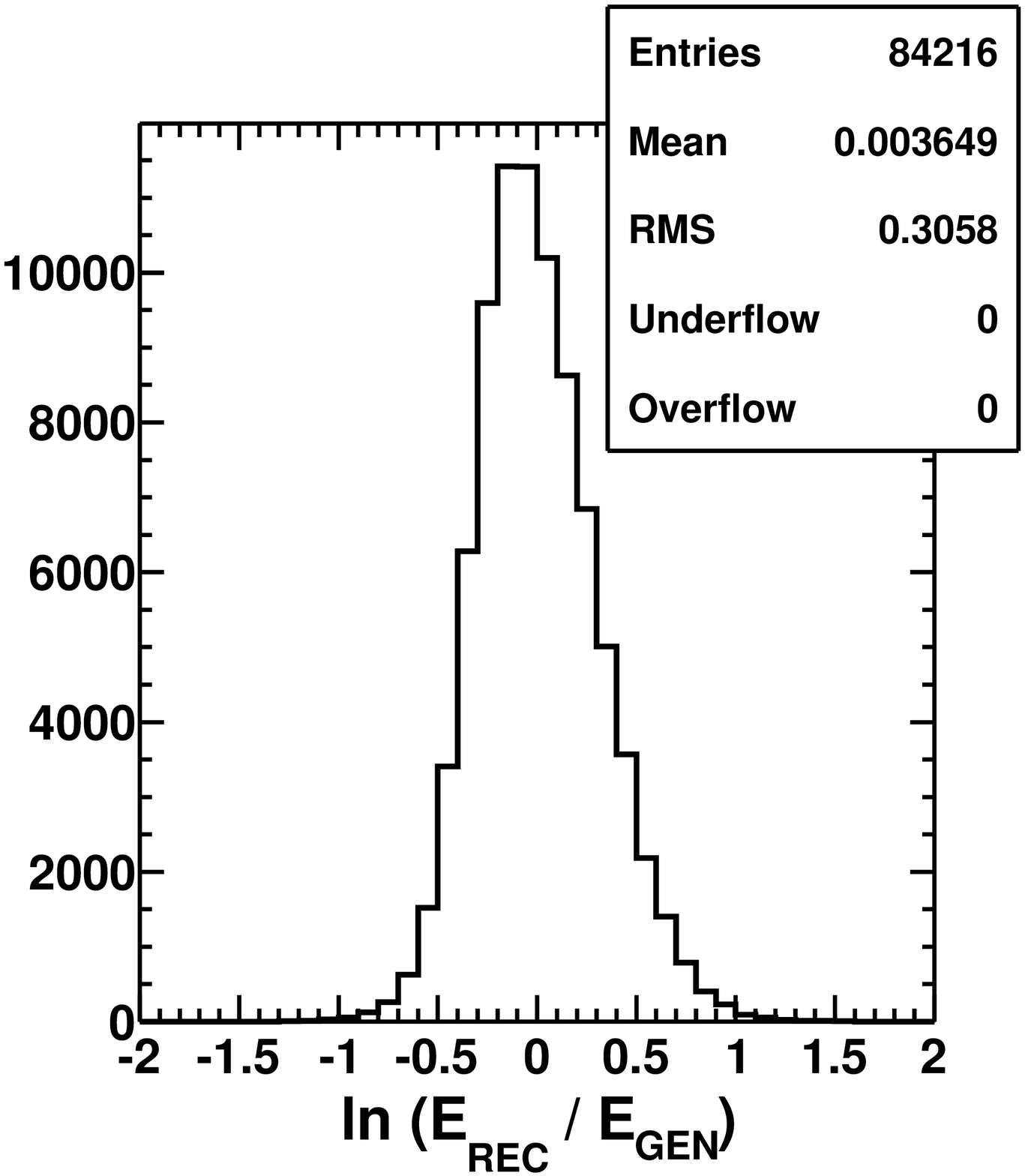}\label{figure:tasd_eres1}}
    \subfloat[]{\includegraphics[width=0.33\textwidth]{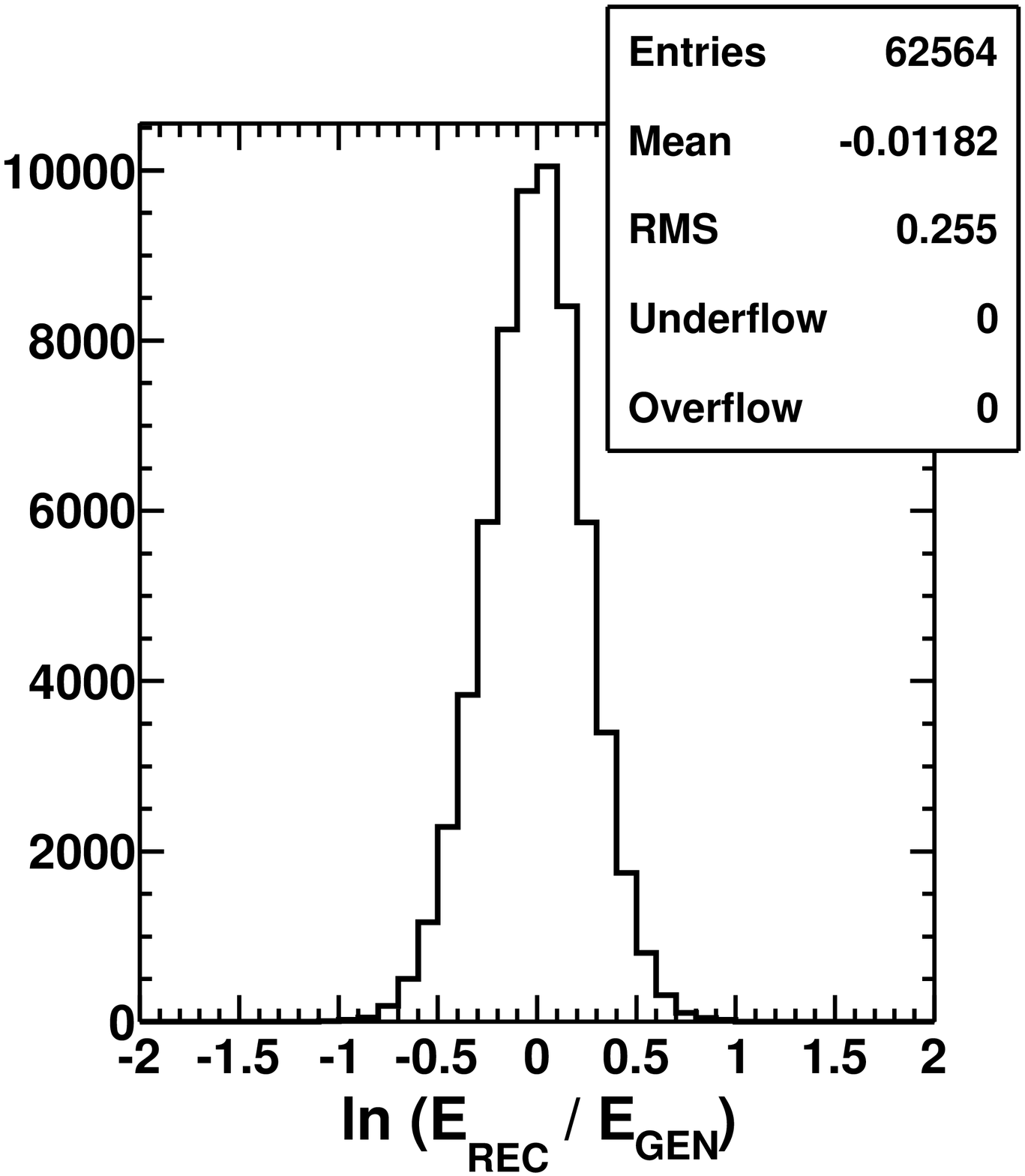}\label{figure:tasd_eres2}}
    \subfloat[]{\includegraphics[width=0.33\textwidth]{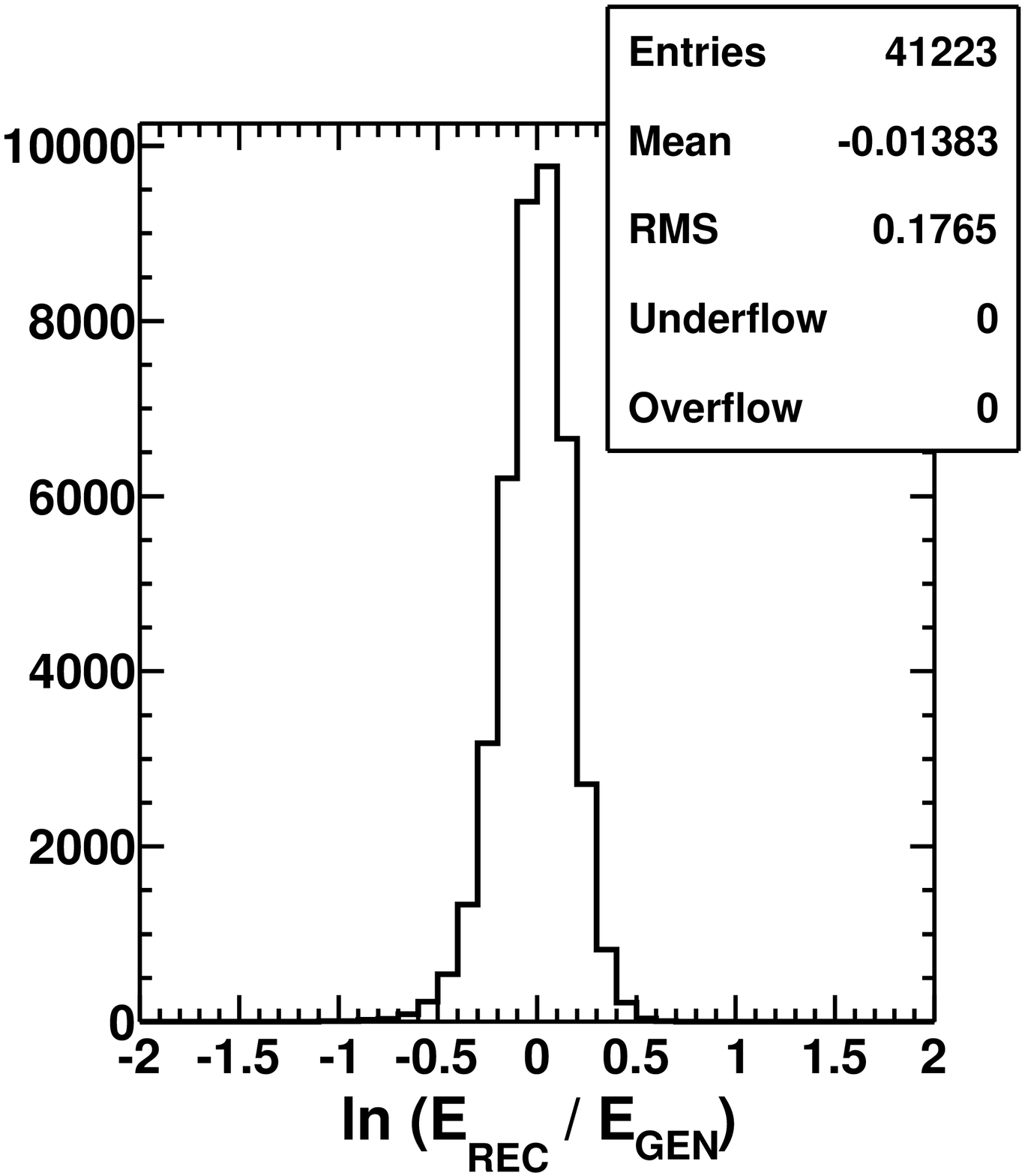}\label{figure:tasd_eres3}}
  }
  \caption{ The TA SD energy resolution evaluated using the
    Monte-Carlo spectral sets.  Energy resolution is shown for three
    ranges in MC generated energy:
    \protect\subref{figure:tasd_eres1} $10^{18.0}\mathrm{eV} <
    E_{\mathrm{GEN}} < 10^{18.5}\mathrm{eV}$,
    \protect\subref{figure:tasd_eres2} $10^{18.5}\mathrm{eV} <
    E_{\mathrm{GEN}} < 10^{19.0}\mathrm{eV}$,
    \protect\subref{figure:tasd_eres3} $E_{\mathrm{GEN}} >
    10^{19.0}\mathrm{eV}$.  Natural logarithm of the ratio         $R=E_{\mathrm{REC}}/E_{\mathrm{GEN}}$ was used for
    producing the histograms. }
  \label{figure:tasd_eres}
\end{figure*}

The RMS deviation $\sigma_{\mathrm{ln}R}$
of the distribution of the (natural) logarithm of the $R=E_{\mathrm{REC}}/E_{\mathrm{GEN}}$
can also be used to calculate $\sigma_{E}$, the fractional energy resolution,
according to the first order approximation:
\begin{equation}
  \label{equation:eEnRes} \sigma_{E} =
\mathrm{exp}(\sigma_{\mathrm{ln}E}) - 1
\end{equation} Figure~\ref{figure:tasd_eres} shows the energy
resolution of the TA SD for three MC generated energy ranges.  The
histograms were produced using the MC spectral sets with varying
statistics (10 to 40 times that of the real data) to yield similar
numbers of events in the histograms.  Using the RMS deviation of the
$E_{\mathrm{REC}}/E_{\mathrm{GEN}}$ distributions and
equation~\ref{equation:eEnRes}, the following results were obtained
for the TA SD energy resolution (in percents of the true energy): 36\%
for $10^{18.0}\mathrm{eV} < E_{\mathrm{GEN}} < 10^{18.5}\mathrm{eV}$,
29\% for $10^{18.5}\mathrm{eV} < E_{\mathrm{GEN}} <
10^{19.0}\mathrm{eV}$, and 19\% for $E_{\mathrm{GEN}} >
10^{19.0}\mathrm{eV}$.

\section{Normalizing the Energy Scale}

\begin{figure*}[t,b]
  \centering 
  \subfloat[SD energy vs. FD energy]{\includegraphics[width=0.5\textwidth]{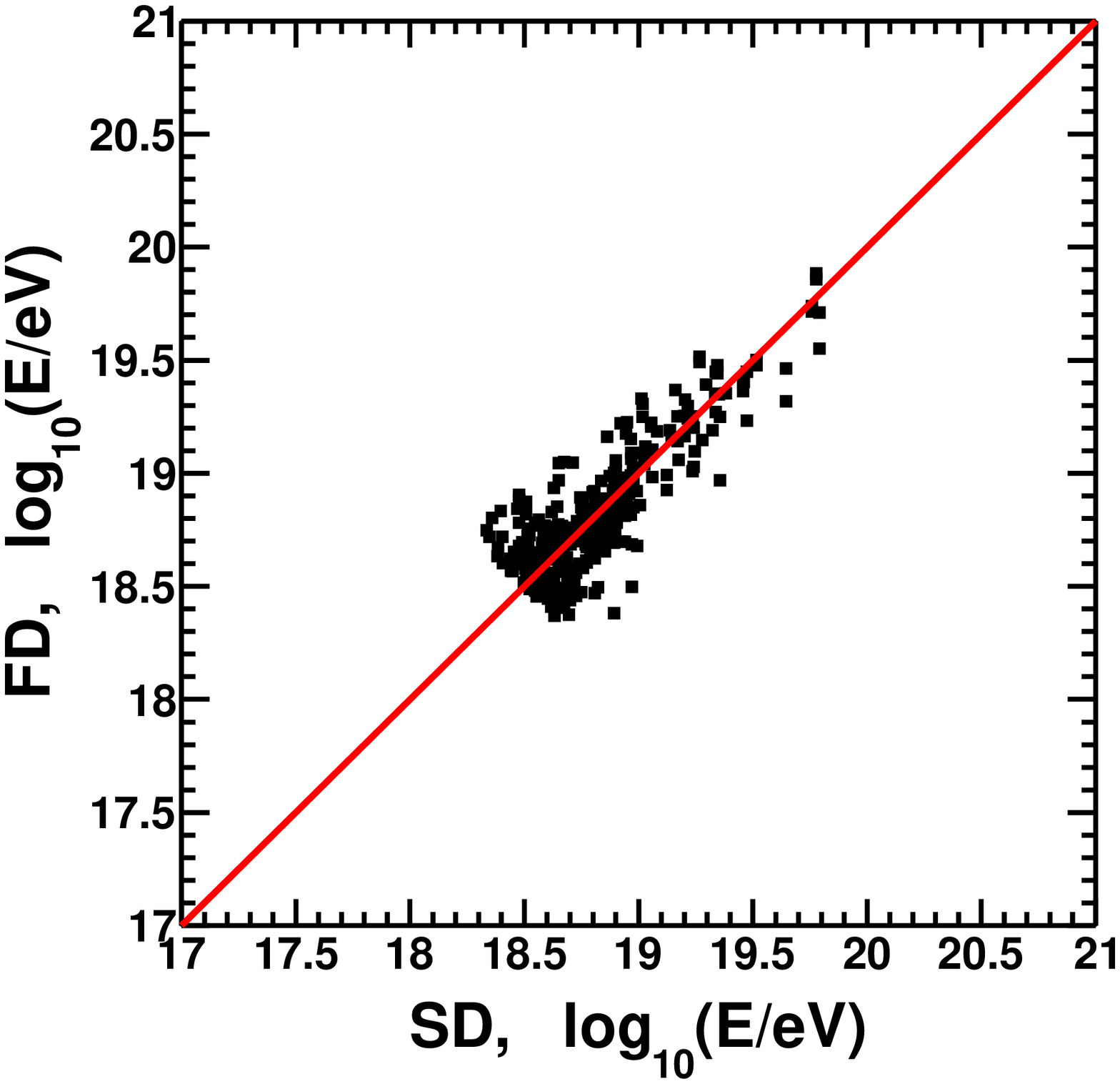}\label{figure:enscale_scatter}}
  \subfloat[SD/FD energy ratio]{\includegraphics[width=0.5\textwidth]{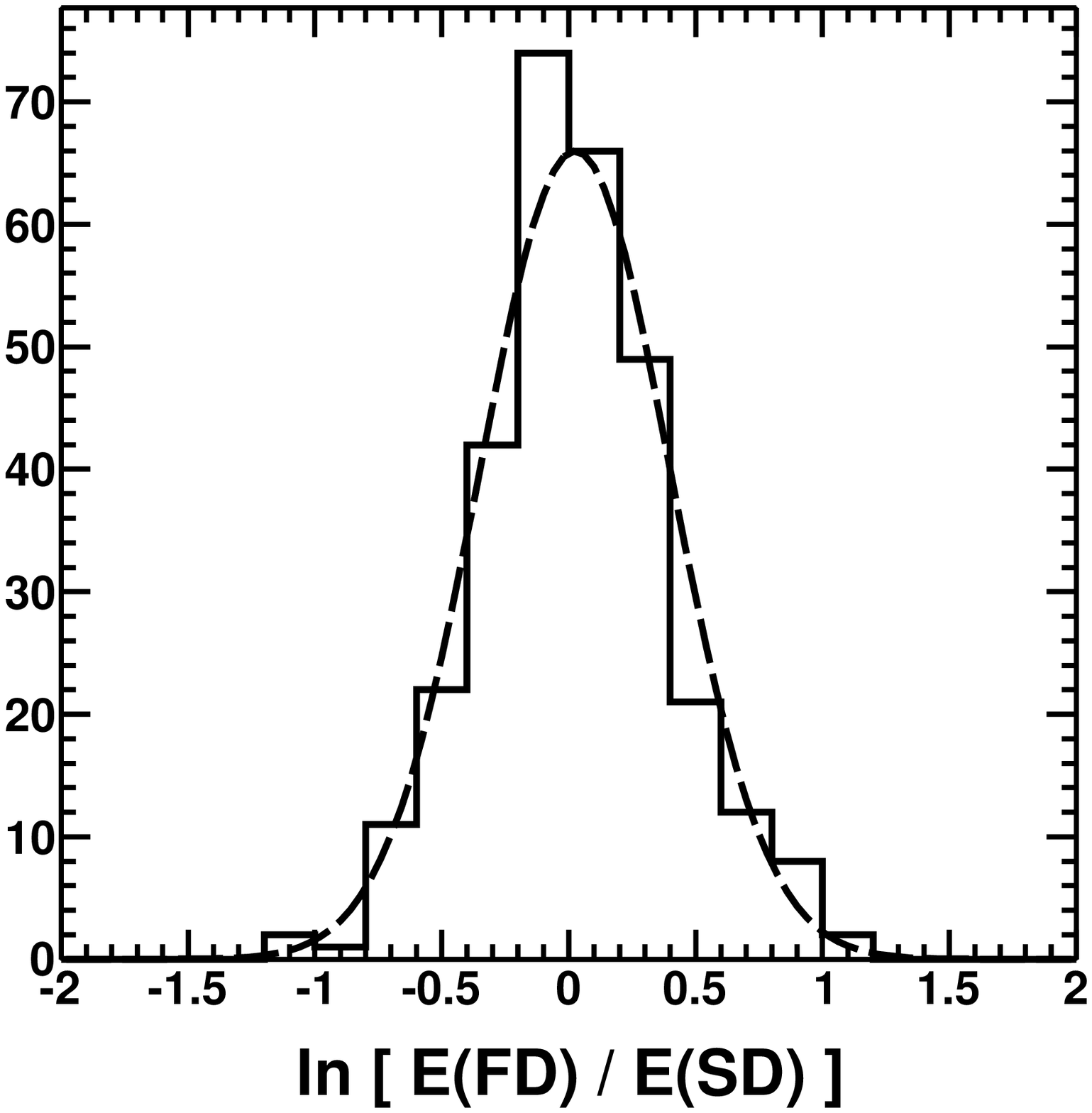}\label{figure:enscale_ratio}}
  \linebreak
  \caption{Energy normalization by FD.  Part (a) shows a scatter plot of SD and FD energies for events seen by both detectors, after the 27\% correction has been applied (see text); and part (b) shows the SD/FD energy ratio after correction.}
  \label{figure:fdsd}
\end{figure*}

The energy of an air shower seen by a fluorescence detector can be measured accurately because the fluorescence process is basically calorimetric.  However for the tails of an air shower, which are observed by a surface detector, one is subject to a much larger uncertainty in energy, which comes from the details of the hadronic generator program used (in our case QGSJET-II).  The size of this uncertainty is unknown.  Therefore a hybrid experiment, like the Telescope Array, has available to it an excellent way of normalizing its SD energy scale:  for events seen by both detectors determine the energy from each detector and normalize the SD energy scale to that of the FD.  We observe a 27\% difference between the two energy scales, which is independent of energy (SD is higher than FD).  We therefore lower the energies of our SD events by this ratio.  Figure~\ref{figure:fdsd} shows a scatter plot of events' energies from the FD and SD after this correction is made.  This normalization is subject to the systematic uncertainty of the TA FD energy scale, which is 22\%~\cite{ta:enscale_sys}.

\section{Study of Systematic Errors}

While calculating the acceptance of the TA surface detector, as is described above, it is possible to estimate the uncertainty in the acceptance values from systematic sources.  In this section we describe such estimates for four sources of systematic uncertainty:  
the attenuation correction used to determine events' energies as a function of $S800$ and zenith angle, changing cuts used to remove poorly reconstructed events, unfolding the SD energy resolution, and the small mismatch between data and Monte Carlo distributions of important quantities.

\subsection{Attenuation of {\rm S}800}
\begin{figure}[t,b]
  \centering
  \includegraphics[width=0.5\textwidth]{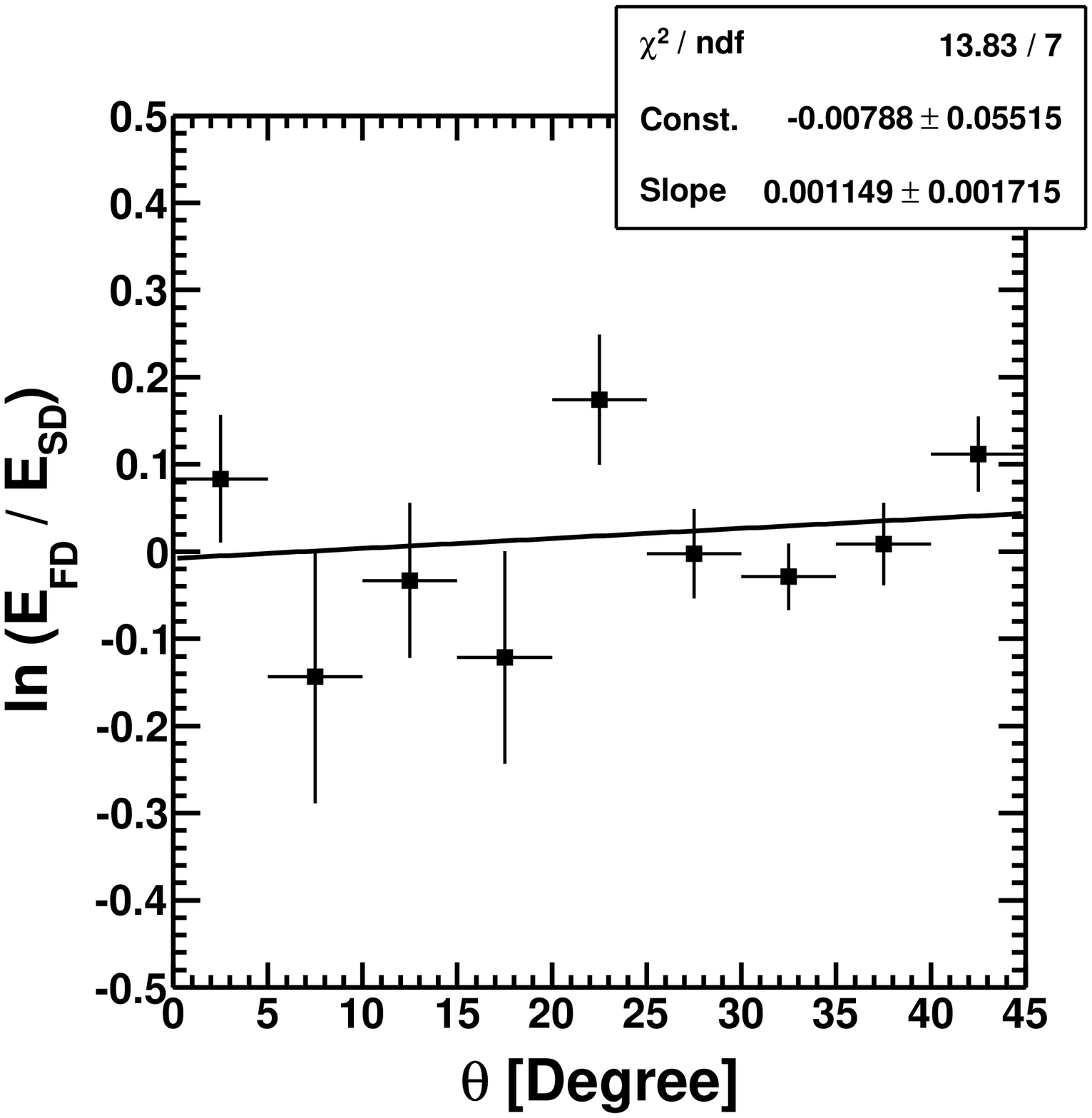}
  \caption { Mean (logarithm) of ratio of the FD and SD energies
    plotted versus the event zenith angle for events well reconstructed by
    the two detectors (points with error bars).  Solid line shows the
    linear fit.  The slope of the fit is within its statistical
    uncertainty.  }
  \label{figure:enscale_vs_theta}
\end{figure}

The $S800$ attenuation correction arises because at different zenith angles
a shower traverses different amounts of atmospheric material before it reaches
the ground and is thus observed at a different stage of shower development.
We check the systematic uncertainties of $S800$ attenuation by comparing
the dependence of the ratio of FD over SD energies plotted versus the
event zenith angle, for events well reconstructed by the FD and SD
(the same events used in producing plots in
Figure~\ref{figure:fdsd}).
Figure~\ref{figure:enscale_vs_theta} shows the result. To measure a possible 
bias in our attenuation correction, we fit the ratios to a straight line.  The
slope of the line is $0.0011\pm0.0017$, showing that no bias can be seen with 
the current statistical power of the data.
\subsection{Acceptance}
\begin{figure*}[t,b]
  \centerline{
    \subfloat[]{\includegraphics[width=0.5\textwidth]
      {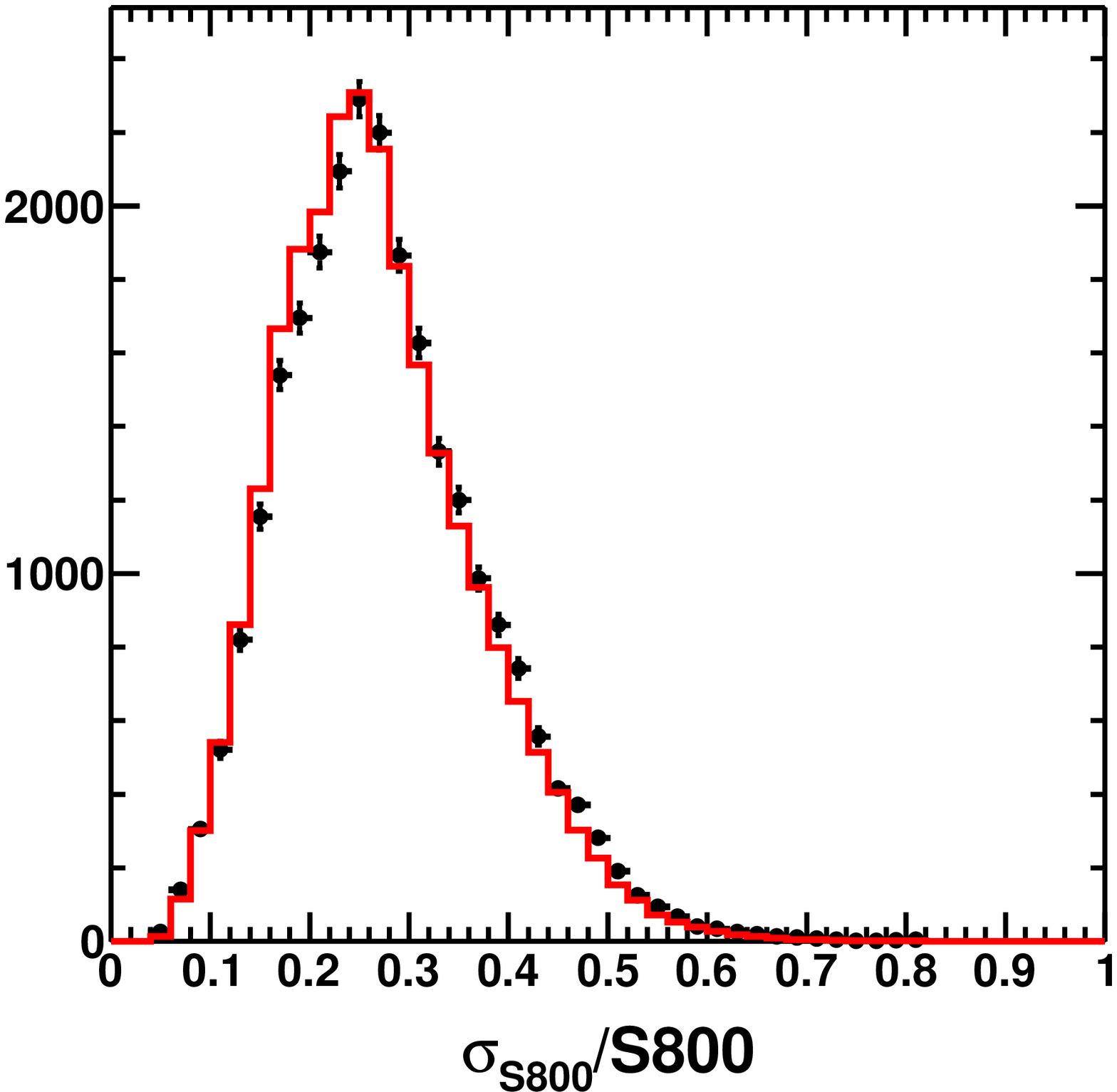}\label{figure:dtmc_sigmaS800_180}}
    \subfloat[]{\includegraphics[width=0.5\textwidth]
      {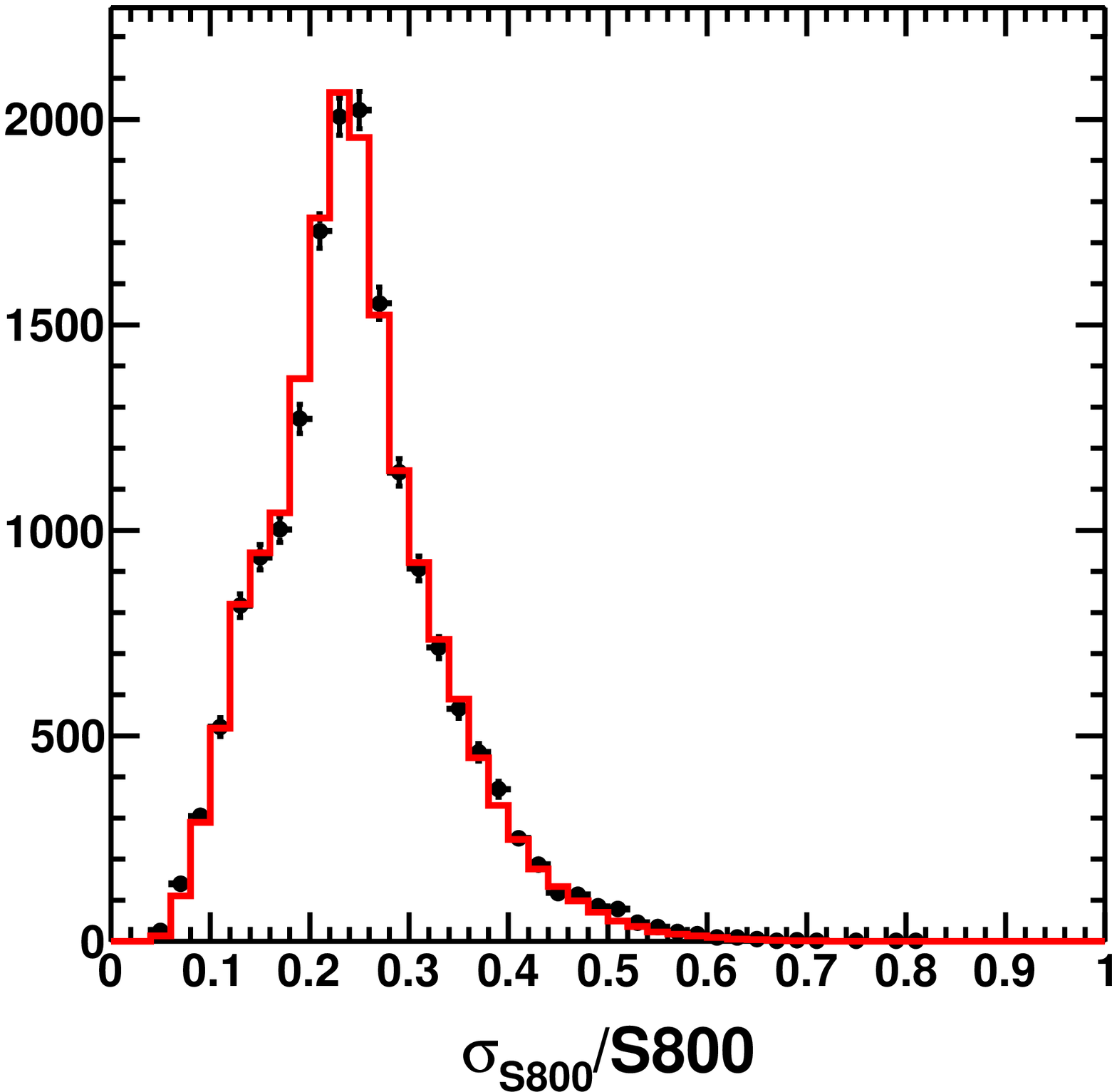}\label{figure:dtmc_sigmaS800_182}}
  }
  \caption { Data and Monte-Carlo comparison of the fraction
    uncertainty on $S800$: \protect\subref{figure:dtmc_sigmaS800_180}: $E >
    10^{18.0}$~eV, \protect\subref{figure:dtmc_sigmaS800_182}: $E >
    10^{18.2}$~eV.  Note the agreement between the data and
    Monte-Carlo becomes better for $E > 10^{18.2}$~eV range.
  }
  \label{figure:dtmc_sigmaS800}
\end{figure*}

An acceptance bias can occur if there are disagreements between the
data and the Monte-Carlo in quantities that are used for making 
quality cuts.  While most quantities agree on a $\sim2\%$ level (the
answer is obtained by counting the differences in the numbers of
events between the data and MC in each bin and dividing that by the
total number of events), there are important exceptions, where the
peaks of the data and MC histograms do not match exactly, i.e. data
and MC histograms are slightly shifted with respect to each other and
simply counting the event differences in bins does not work.  One such
quantity is the fractional uncertainty in
$S800$, shown in Figure~\ref{figure:dtmc_sigmaS800}.

This quantity is used for making the quality cuts:
$\sigma_{S800}/S800 < 0.25$.  To
determine the systematic uncertainty (effect on the flux) due to
this cut, we consider the data and Monte-Carlo ratio:
\begin{equation}
  R_{i} = \frac{(N^{\mathrm{DATA}}_{\mathrm{REC}})_{i}}{(N^{\mathrm{MC}}_{\mathrm{REC}})_{i}} \,,
  \label{equation:dtmc_ratio}
\end{equation}
where $(N^{\mathrm{DATA}}_{\mathrm{REC}})_{i}$and
$(N^{\mathrm{MC}}_{\mathrm{REC}})_{i}$ are numbers of events
reconstructing in the $i^{th}$ ($\mathrm{log}_{10}E$) energy bin for
data and Monte-Carlo, respectively.  The systematic uncertainty can
be readily estimated by calculating $R_{i}$ with the cut
($R_{i}^{\mathrm{CUT}}$) and without the cut
($R_{i}^{\mathrm{NO\,CUT}}$) and evaluating the fractional difference:
\begin{equation}
  B_{i} = R_{i}^{\mathrm{WITH\,CUT}} / R_{i}^{\mathrm{NO\,CUT}} - 1
  \label{equation:acceptance_bias}
\end{equation}

\begin{figure*}[t,b]
  \centering
    \includegraphics[width=0.5\textwidth] {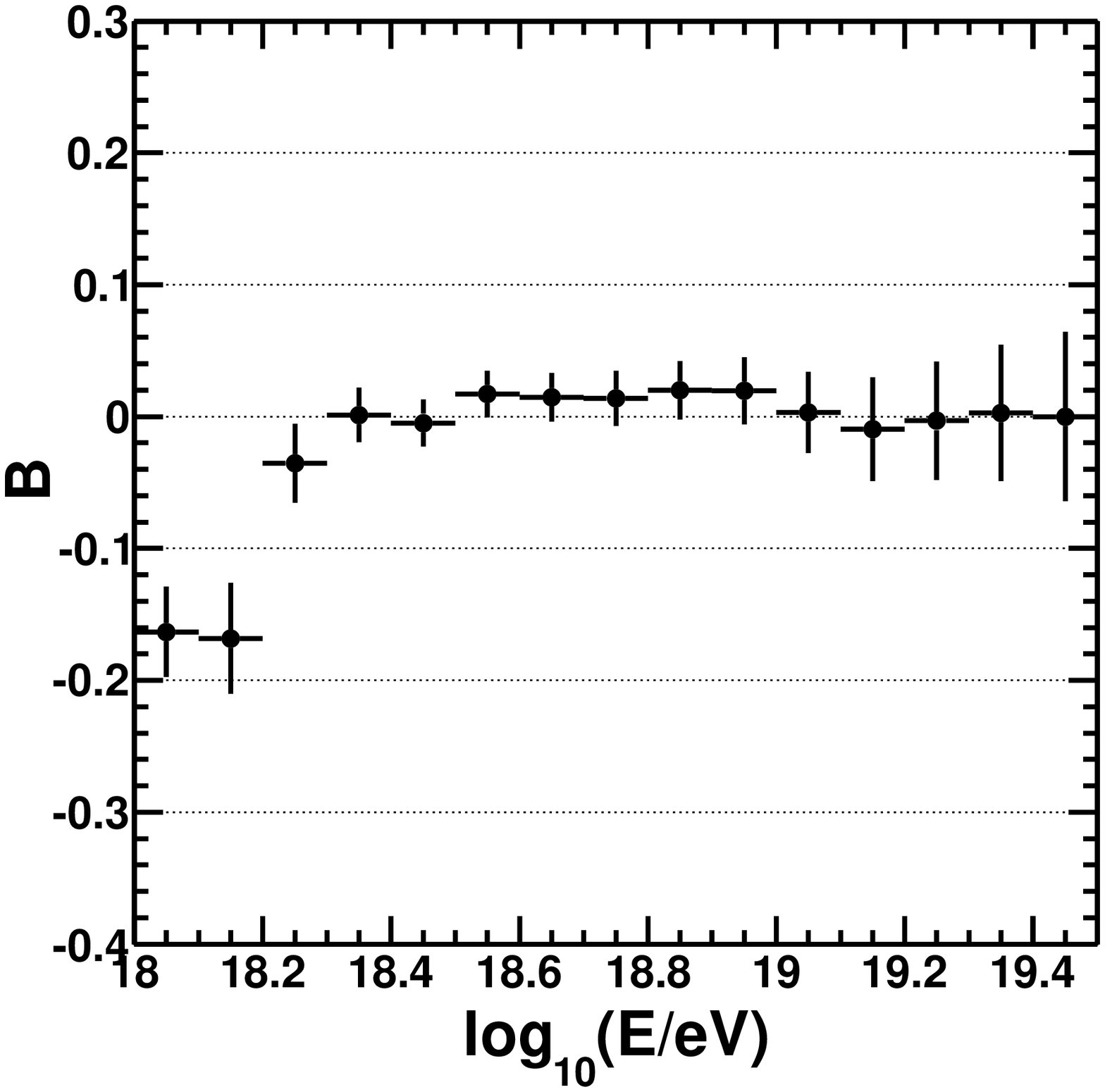}\label{figure:sigmaS800_bias}
  \caption {Fractional change in the flux after eliminating the
    $\sigma_{S800}/S800 < 0.25$  cut
    (estimated using Equation~\ref{equation:acceptance_bias}) plotted
    versus the (reconstructed) event energy.  For $E > 10^{18.2}$~eV,
    the variation is within
    $\sim2\%$.  }
  \label{figure:acceptance_bias}
\end{figure*}

Figure~\ref{figure:acceptance_bias} shows the bias, $B_{i}$, evaluated 
for the cut on $\sigma_{S800}/S800$.  
It shows the systematic change due to the
$\sigma_{S800}/S800$ cut is $\sim2\%$ for $E > 10^{18.2}$~eV.  If one chooses energies $>10^{18.2}$~eV for calculating the TA SD
spectrum, one can avoid any bias as shown in Figure~\ref{figure:acceptance_bias}.

\subsection{Resolution Unfolding}

In calculating the energy spectrum, the resolution of the detector (especially if it is non-Gaussian), coupled with a spectrum that rapidly falls with energy, can bias the result.  Consequently, we must make a first-order resolution correction in the spectrum calculation.  The formula we use is:  
\begin{equation}
J(E) = \frac{T(E')}{A(E)} \frac{D(E)}{A\Omega\Delta t},
\label{eq:spec}
\end{equation} 
where $J(E)$ is the flux, $D(E)$ is the number of events observed in an energy bin, $A(E)$ is the number of accepted simulated events in the energy bin, and $T(E')$ is the number of thrown simulated events.  The surface area, solid angle acceptance, and live time of the detector are repesented by $A, \Omega$, and $\Delta t$, respectively.  Finally, $E$ is the reconstructed energy, while $E'$ is the thrown energy.  

In the case of an ideal detector with perfect resolution
and 100\% efficiency, $T(E')/A(E)=1$ and $J(E)=D(E)/(A\Omega\Delta t)$.  For a
real detector with finite resolution and less than perfect efficiency, the ratio
$T(E')/A(E)$ performs two important roles.  First, $T/A$ compensates for detector efficiency.  Second, by binning $A$ in $E$ but $T$ in $E'$, we perform a first
order (bin-by-bin) correction for energy resolution.  
The validity of this correction is contigent upon energy resolution 
that is the same size or smaller than the energy binning used in the 
calculation,
accurate simulation of the energy resolution (as established by the excellent agreement between data and simulation for the $\chi^{2}/dof$ of the lateral distribution fits in Figure~\ref{figure:dtmc_ldfchi2pdof}), and the use of a 
reasonably accurate input spectral index for the simulation.

If there is a sharp bend in the spectrum, for example the GZK cutoff, the spectrum put into the Monte Carlo should also have the sharp bend, to achieve the best accuracy. While we did not include the effect in the MC for the Data-MC comparison earlier, the GZK cut-of, as previously observed by HiRes, was included in the aperture calculation for the spectrum measurement.  The result for the TA SD spectrum is a level of resolution-generated bias that is much smaller than the statistical power of the experiment.

\subsection{Uncertainty in Energy Scale and Flux}

The systematic uncertainty $\sigma_{J}^{\mathrm{SYS},E}$ on the flux
$J$ due to the systematic uncertainty of the energy scale
$\sigma_{E}^{\mathrm{SYS}}$ can be estimated as follows:
\begin{equation}
  \frac{ \sigma_{J}^{\mathrm{SYS},E} }{ J } = 
  \frac{ \sigma_{N}^{\mathrm{SYS},E} }{ N } = 
  \frac{\Big|\mathrm{d}N/\mathrm{d}E\Big|}{N} \sigma_{E}^{\mathrm{SYS}} = 
  (\gamma-1)\,\frac{\sigma_{E}^{\mathrm{SYS}}}{E} \, ,
  \label{equation:escale_sys}
\end{equation}
where $\gamma$ is the measured spectral index.  The spectral index for
the TA SD spectrum above the ankle is taken from the publication
describing the measurement \cite{sdspec}: $\gamma \simeq 2.67$ and the
systematic uncertainty of the energy scale is controlled by the TA FD:
$\sigma_{E}^{\mathrm{SYS}} / E \simeq 22\%$ \cite{ta:enscale_sys}.
Together, these results give $\sigma_{J}^{\mathrm{SYS},E} / J \simeq
37\%$.  The fluorescence energy scale uncertainty dominates the systematic 
uncertainty 
in energy at $\pm 22\%$.  All other contributions explored here change the 
answer by $<1\%$.  Thus the total systematic uncertainty in $J$ is 37\%.

\section{Conclusion}

We have demonstrated that the dethinned CORSIKA/QGSJET-II-03 proton Monte Carlo simulation
accurately models the response of the TA array of scintillation counters
to cosmic rays in the $E>10^{18.2}$eV, $\theta < 45^{o}$ domain. 
In reconstruction of events, fits to counter times and pulse heights are almost identical for the Monte Carlo
and the data.  Basic histograms of geometrical quantities, and of quantities related to the lateral distribution
of counter pulse heights, for the Monte Carlo agree very well with the same distributions
for the data.  We have measured a 27\% correction to the energy scale of the CORSIKA + QGSJET-II simulation package based on air showers observed calorimetrically by the Telescope Array fluorescence detector, and examined some sources of systematic errors in our aperture calculation.  We conclude that this Monte Carlo simulation is an accurate tool for calculating the surface detector
aperture used to calculate the energy spectrum, as well as to estimate the exposure on the sky for cosmic ray anisotropy analyses.

\section{Acknowledgments}

The Telescope Array experiment is supported 
by the Japan Society for the Promotion of Science through
Grants-in-Aid for Scientific Research on Specially Promoted Research (21000002) 
``Extreme Phenomena in the Universe Explored by Highest Energy Cosmic Rays'', 
and the Inter-University Research Program of the Institute for Cosmic Ray 
Research;
by the U.S. National Science Foundation awards PHY-0307098, 
PHY-0601915, PHY-0703893, PHY-0758342, and PHY-0848320 (Utah) and 
PHY-0649681 (Rutgers); 
by the National Research Foundation of Korea 
(2006-0050031, 2007-0056005, 2007-0093860, 2010-0011378, 2010-0028071, R32-10130);
by the Russian Academy of Sciences, RFBR
grants 10-02-01406a and 11-02-01528a (INR),
IISN project No. 4.4509.10 and 
Belgian Science Policy under IUAP VI/11 (ULB).
The foundations of Dr. Ezekiel R. and Edna Wattis Dumke,
Willard L. Eccles and the George S. and Dolores Dore Eccles
all helped with generous donations. 
The State of Utah supported the project through its Economic Development
Board, and the University of Utah through the 
Office of the Vice President for Research. 
The experimental site became available through the cooperation of the 
Utah School and Institutional Trust Lands Administration (SITLA), 
U.S.~Bureau of Land Management and the U.S.~Air Force. 
We also wish to thank the people and the officials of Millard County,
Utah, for their steadfast and warm support. 
We gratefully acknowledge the contributions from the technical staffs of our
home institutions and the University of Utah Center for High Performance Computing
(CHPC). 


\newpage
\bibliographystyle{elsarticle-num}
\bibliography{sddtmc}

\end{document}